\documentclass[sigconf, nonacm]{acmart}

\newcommand\vldbdoi{XX.XX/XXX.XX}
\newcommand\vldbpages{XXX-XXX}
\newcommand\vldbvolume{17}
\newcommand\vldbissue{2}
\newcommand\vldbyear{2023}
\newcommand\vldbauthors{\authors}
\newcommand\vldbtitle{\shorttitle} 
\newcommand\vldbavailabilityurl{URL_TO_YOUR_ARTIFACTS}
\newcommand\vldbpagestyle{plain} 

\usepackage{listings}

\definecolor{eclipseStrings}{RGB}{42,0.0,255}
\definecolor{eclipseKeywords}{RGB}{127,0,85}
\colorlet{numb}{magenta!60!black}

\lstdefinelanguage{json}{
    basicstyle=\normalfont\ttfamily,
    commentstyle=\color{eclipseStrings}, 
    stringstyle=\color{eclipseKeywords}, 
    numbers=left,
    numberstyle=\scriptsize,
    stepnumber=1,
    numbersep=8pt,
    showstringspaces=false,
    breaklines=true,
    frame=lines,
    string=[s]{"}{"},
    comment=[l]{:\ "},
    morecomment=[l]{:"},
    literate=
        *{0}{{{\color{numb}0}}}{1}
         {1}{{{\color{numb}1}}}{1}
         {2}{{{\color{numb}2}}}{1}
         {3}{{{\color{numb}3}}}{1}
         {4}{{{\color{numb}4}}}{1}
         {5}{{{\color{numb}5}}}{1}
         {6}{{{\color{numb}6}}}{1}
         {7}{{{\color{numb}7}}}{1}
         {8}{{{\color{numb}8}}}{1}
         {9}{{{\color{numb}9}}}{1}
}

\usepackage{enumitem}
\setlist{nosep}

\usepackage{fancyhdr}
\usepackage[normalem]{ulem}
\usepackage[noend]{algpseudocode}
\usepackage{algorithm}
\usepackage{tcolorbox}
\usepackage{tikz}
\usepackage{amsfonts}
\usepackage{eucal}

\usepackage{blindtext,graphicx}
\usepackage[absolute]{textpos}
\newcommand{\textllsm}[1]{{\fontfamily{lmss}\selectfont #1}}

\usepackage[compact]{titlesec}
 \titlespacing*{\section}{0pt}{3pt}{3pt}
 \titlespacing*{\subsection}{0pt}{1pt}{1pt}
 
\newcommand{\THISWORK}{{\fontfamily{lmss}\selectfont
Everest}}
\newcommand*\circled[1]{\tikz[baseline=(char.base)]{\node[shape=circle,draw,inner sep=0.85pt, fill=black] (char) {\textcolor{white}{#1}};}}

\newcommand{\WIKI}{{\fontfamily{lmss}\selectfont
wiki-talk}}

\newcommand{\ETH}{{\fontfamily{lmss}\selectfont
ethereum}}

\newcommand{\linebreakand}{%
  \end{@IEEEauthorhalign}
  \hfill\mbox{}\par
  \mbox{}\hfill\begin{@IEEEauthorhalign}
}

\begin{document}
\begin{textblock}{16}(3,1)
{\normalsize \normalfont \textit{Authors’ version; to appear in the Proceedings of 50th International Conference on Very Large Databases (VLDB 2024).} }
\end{textblock}

\title{\THISWORK: GPU-Accelerated System For Mining Temporal Motifs
}

\author{Yichao Yuan}
\affiliation{%
  \institution{University of Michigan}
  \city{Ann Arbor, Michigan}
  \state{USA}
}
\email{yichaoy@umich.edu}

\author{Haojie Ye}
\affiliation{%
  \institution{University of Michigan}
  \city{Ann Arbor, Michigan}
  \state{USA}
}
\email{yehaojie@umich.edu}

\author{Sanketh Vedula}
\affiliation{%
  \institution{Technion}
  \city{Haifa}
  \country{Israel}
}
\email{sanketh@campus.technion.ac.il}

\author{Wynn Kaza}
\affiliation{%
  \institution{University of Michigan}
  \city{Ann Arbor, Michigan}
  \state{USA}
}
\email{wynnkaza@umich.edu}

\author{Nishil Talati}
\affiliation{%
  \institution{University of Michigan}
  \city{Ann Arbor, Michigan}
  \state{USA}
}
\email{talatin@umich.edu}


\makeatletter



\begin{abstract}
Temporal motif mining is the task of finding the occurrences of subgraph patterns within a large input temporal graph that obey the specified structural and temporal constraints.
Despite its utility in several critical application domains that demand high performance (\textit{e.g.,} detecting fraud in financial transaction graphs), the performance of existing software is limited on commercial hardware platforms, in that it runs for tens of hours.
This paper presents \THISWORK---a system that efficiently maps the workload of mining (supports both enumeration and counting) temporal motifs to the highly parallel GPU architecture.
In particular, using an input temporal graph and a more expressive user-defined temporal motif query definition compared to prior works, \THISWORK\ generates an execution plan and runtime primitives that optimize the workload execution by exploiting the high compute throughput of a GPU.
\THISWORK\ generates motif-specific mining code to reduce long-latency memory accesses and frequent thread divergence operations.
\THISWORK\ incorporates novel low-cost runtime mechanisms to enable load balancing to improve GPU hardware utilization.
To support large graphs that do not fit on GPU memory, \THISWORK\ also supports multi-GPU execution by intelligently partitioning the edge list that prevents inter-GPU communication.
\THISWORK\ hides the implementation complexity of presented optimizations away from the targeted system user for better usability.
Our evaluation shows that, using proposed optimizations, \THISWORK\ improves the performance of a baseline GPU implementation by 19$\times$, on average.
\end{abstract}

\maketitle

\pagestyle{\vldbpagestyle}
\begingroup\small\noindent\raggedright\textbf{PVLDB Reference Format:}\\
\vldbauthors. \vldbtitle. PVLDB, \vldbvolume(\vldbissue): \vldbpages, \vldbyear.\\
\href{https://doi.org/\vldbdoi}{doi:\vldbdoi}
\endgroup
\begingroup
\renewcommand\thefootnote{}\footnote{\noindent
This work is licensed under the Creative Commons BY-NC-ND 4.0 International License. Visit \url{https://creativecommons.org/licenses/by-nc-nd/4.0/} to view a copy of this license. For any use beyond those covered by this license, obtain permission by emailing \href{mailto:info@vldb.org}{info@vldb.org}. Copyright is held by the owner/author(s). Publication rights licensed to the VLDB Endowment. \\
\raggedright Proceedings of the VLDB Endowment, Vol. \vldbvolume, No. \vldbissue\ %
ISSN 2150-8097. \\
\href{https://doi.org/\vldbdoi}{doi:\vldbdoi} \\
}\addtocounter{footnote}{-1}\endgroup

\ifdefempty{\vldbavailabilityurl}{}{
\vspace{.3cm}
\begingroup\small\noindent\raggedright\textbf{PVLDB Artifact Availability:}\\
The source code, data, and/or other artifacts have been made available at \url{https://github.com/yichao-yuan-99/Everest}.
\endgroup
}

\section{Introduction}
Graph (or network) is a ubiquitous data structure that models various types of entities and their interactions through the collection of vertices and edges.
For example, graphs can model user interactions in social and communication networks~\cite{kovanen2011temporal, kovanen2013temporal, lahiri2007structure, paranjape2017motifs}, transactions in financial networks~\cite{hajdu2020temporal}, and protein/drug interactions in biological networks~\cite{meydan2013prediction}.
Small subgraph patterns called \textit{motifs} enable the understanding of the structure and function of complex systems encoded in real-world graphs~\cite{shen2002network, alon2007network,newman2003structure}.
Mining motifs within large graphs is one of the fundamental problems in network science~\cite{milo2002network}.

Most real-world phenomena are dynamic in nature.
Static graphs aggregate the interactions that occur over the graph by omitting the temporal information, leading to information loss and performance degradation in downstream tasks.
Temporal graph is a type of dynamic graph that stores network interaction information in terms of timestamped edges between nodes.
Therefore, temporal graphs capture richer information compared to static networks~\cite{kovanen2011temporal, pan2011path}.
\textit{Temporal motif} is a fundamental building block of temporal graphs, similar to static motif for static graphs.
Temporal motifs are useful for several critical and main-stream applications including financial fraud detection~\cite{hajdu2020temporal}, insider threat identification in an organization~\cite{glasser2013bridging, mackey2018chronological}, energy disaggregation monitoring on electrical grids~\cite{shao2013temporal}, and peptide binding prediction in structural biology~\cite{meydan2013prediction}.
Furthermore, motif counts have been shown to resolve symmetries and improve the expressive power of graph neural networks~\cite{bouritsas2022improving,chamberlain2022graph}. 

The wide applicability and criticality of application domains necessitate high performance of temporal motif mining.
However, existing software solutions~\cite{paranjape2017motifs,mackey2018chronological,kumar20182scent,wang2020efficient,liu2018sampling,sarpe2021presto,kovanen2011temporal,pasha21Kdd} offer sub-optimal performance on CPU platforms.
This is because of the high computation complexity of mining temporal motifs.
While GPU is an attractive solution to accelerate this workload, our study shows that a straightforward GPU implementation based on the state-of-the-art temporal motif mining algorithm~\cite{mackey2018chronological} heavily under-utilizes GPU resources.
For example, our experiments show that mining a temporal motif four edges in a graph with about 600 million temporal edges takes more than 18 GPU hours.
This is due to long-latency memory operations, frequent thread divergence due to excessive branching code, and heavy load imbalance inherent to this workload.
Furthermore, these solutions are limited in terms of their expressive power representing temporal motif queries.

This paper presents \THISWORK---an optimized system to mine temporal motifs on the GPU architecture.
The inputs to \THISWORK\ are 1) temporal graph (a type of dynamic property graph), and 2) user-defined query that specifies a temporal motif to mine within the input graph.
\THISWORK\ supports an expressive definition of an input query, where a user can specify (a) the motif structure/connectivity pattern, (b) temporal edge order, (c) both coarse (between the last and first edge~\cite{mackey2018chronological,paranjape2017motifs}) and fine-grained (between consecutive edges~\cite{kovanen2011temporal,pasha21Kdd}) $\delta$-temporal window constraints, (d) constraints on vertex/edge attributes, as well as (e) temporal anti-edge(s) (\textit{i.e,} absence of edge(s)).
Using these inputs, \THISWORK\ produces an optimized execution plan and runtime that maps efficiently on the GPU architecture.
For scalability, we also support multi-GPU execution out--of--the--box.
The outout of \THISWORK\ includes exact motif matches, supporting both enumeration and counting.

To optimize the workload execution on the GPU architecture, \THISWORK\ employs the following optimizations.
First, \THISWORK\ extends each GPU thread's context to \textit{cache key metadata information} (\textit{e.g.,} possible motif-graph edge matches) that reduces the number of costly binary search operations that incur long-latency memory accesses.
Second, \THISWORK\ generates \textit{motif-specific mining code} with auxiliary data structures and a modified algorithm based on pre-decoding of the temporal motif structure and temporal constraints.
This reduces the amount of branching code and thread divergence, making the generated software GPU architecture-friendly.

We further introduce \textit{sub-tree-level parallelism} that enables the exploration of different sub-trees within the same search tree concurrently.
Using this concept, third, we introduce \textit{intra-warp work stealing}---a low-cost mechanism that permits different threads in the same warp to steal work from each other.
Fourth, we propose \textit{tail warp work redistribution} that reshuffles the work assignments to different warps when a subset of warps are stuck processing large volumes of work while others are idle.
\THISWORK\ runtime employs these two techniques to perform load balancing for improved hardware utilization.
\THISWORK\ employs the aforementioned optimization "under-the-hood" by hiding their implementation complexity within the system itself; a user only needs to specify a query to mine a specific temporal motif.
To scale workload execution to multiple GPUs, \THISWORK\ employs a low-cost edge list partitioning scheme that completely prevents inter-GPU communication.

To evaluate the performance of \THISWORK, we use 13 unique input temporal motifs and four large-scale temporal graph datasets.
Compared to a baseline CPU and GPU implementations, the proposed optimizations improve the performance by 62.1$\times$ and 19$\times$, on average.
The high performance of \THISWORK\ is attributed to improved compute and memory throughput.
We also show that \THISWORK\ offers near-linear performance scaling for a multi-GPU system for workloads with long execution times.
\THISWORK\ is the \textbf{first work} that designs a domain-specific system that efficiently maps the workload of temporal motif mining to the GPU architecture.
The key contributions of this work are as follows.
\begin{itemize}[leftmargin=*]
    \item \textit{Characterization of challenges} to efficiently execute the workload of temporal motif mining on the GPU architecture.
    \item \textit{Generalized motif query} interface that supports fine-grained temporal constraints, vertex/edge labels, and temporal anti-edges.
    \item \textit{Data structure and algorithmic optimizations} to reduce operations causing long memory latency and frequent thread divergence.
    \item \textit{Runtime load balancing} techniques that redistribute work among different GPU threads and warps to improve hardware utilization.
    \item \textit{\THISWORK:} an end-to-end open-source system that enables high performance by hiding all the implementation complexities from the user to provide 19$\times$ performance uplift compared to a baseline GPU implementation.
\end{itemize}

\section{Background}

\subsection{Problem definition}
\label{subsec:problem_def}
\textbf{Temporal Graph.} We define a \textit{temporal graph} as an ordered set of temporal edges, and a \textit{temporal edge} as a timestamped directed edge between an ordered pair of nodes. 
Explicitly, a temporal graph $G$ is represented as a set of ordered tuples $G = \{ (u_i, v_i, t_i)\}_{i=1}^{m}$, with $u_i$ and $v_i$ denoting the source and destination nodes of the temporal edge $(u_i, v_i)$, respectively, and $t_i \in \mathbb{R}^+$ being the timestamp of the edge. We assume that the edges in $G$ are temporally ordered and have unique timestamps. Furthermore, nodes and edges can be optionally endowed with discrete/continuous attributes/labels.

\noindent \textbf{$\delta$-Temporal Motif.} A $\delta$-temporal motif is defined as a sequence of $l$ edges, $M=\{ (u_i, v_i, t_i) \}_{i=1}^{l}$ that are temporally ordered and occur within a time $\delta \in \mathrm{R}^+$, \textit{i.e.,} $t_1 < t_2 \ldots < t_l$ and $t_l - t_1 \leq \delta$.
Prior works define temporal motif mining as the problem of mining occurrences of the \textit{$\delta$-temporal motif} $M$ within a larger temporal graph $G$.
\THISWORK\ also supports other constraints as detailed below.

\noindent \textbf{Fine-Grained $\delta$-Temporal Motif.} A \textit{fine-grained} $\delta${-temporal motif} is a $\delta$-temporal motif with additional temporal constraints between consecutive edges, $t_{i+1} - t_i \leq \delta_{i}$ where $ \delta_{i} $ is the maximum delay between edges $i$ and $i+1$. Notice that a $\delta$-temporal motif is a fine-grained $\delta$-temporal motif with $\delta_{i} \geq t_{i+1} - t_{i}, \forall i$; thus, the notion of fine-grained $\delta$-temporal motif subsumes $\delta$-temporal motif.

\noindent \textbf{Temporal Anti-Edges~\cite{kovanen2011temporal, pasha21Kdd}.} An edge $e_i = (u_i, v_i, t_i) \in M$ can be ``attached'' with a \textit{temporal anti-edge} $\neg e_{ij} = \neg (u_j, v_j, \delta_{ij})$ implying that we are matching for the \textit{absence} of $e_j = (u_j, v_j, t_j)$ where $t_j \in [t_i, t_i + \delta_{ij}]$.
Intuitively, a temporal edge defines the \textit{inclusion criteria}, versus a temporal anti-edge defines the \textit{exclusion criteria}.

\textbf{Generalized Temporal Motif Mining and Use Case.}
We define \textit{generalized temporal motif mining} as the problem of mining occurrences of a fine-grained $\delta$-temporal motif with optionally added anti-edges and constraints specified over the node and edge attributes.
Typically there are two flavors of output: the matched motifs can either be enumerated or counted.
Our problem definition targets the general demands of finding isomorphism-based motifs, which seek bijections between motif patterns and sub-graphs in the data.
While using a similar structural and temporal definition, few other works~\cite{Kosyfaki2018FlowMI} adopt different semantics for their motifs from isomorphism (\textit{e.g.,} matching motifs edges to sets of graph edges), which is out of the scope of this paper.

\begin{figure}[t]
\centerline{\includegraphics[width=\linewidth]{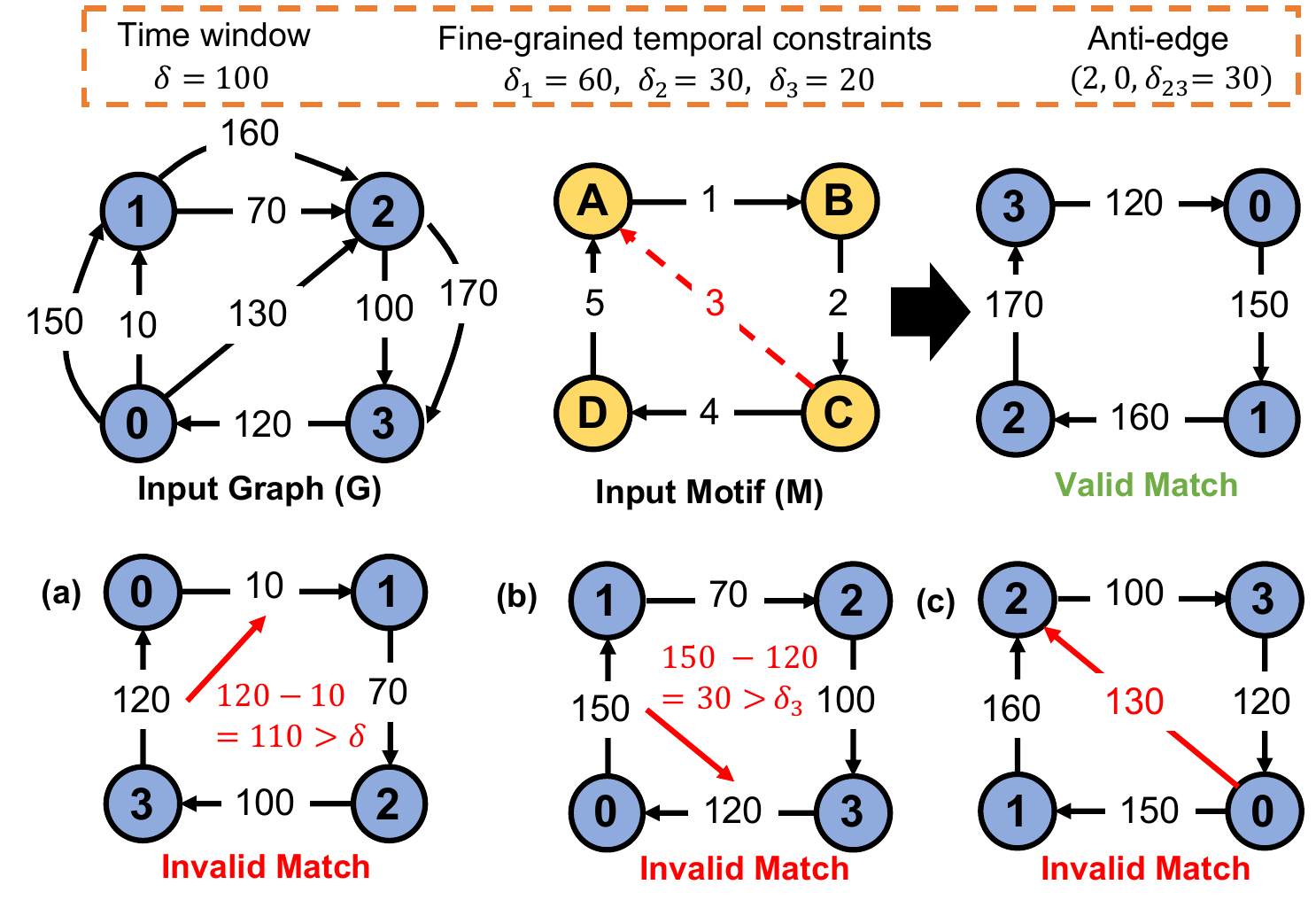}} 
\vspace{-0.5cm}
\caption{A walk-through example of mining an input temporal motif (M) within an input graph (G).}
\vspace{-0.6cm}
\label{fig:walkthrough_example}
\end{figure}
\textbf{A Walk-Through Example.}
Fig.~\ref{fig:walkthrough_example} presents an example that captures different aspects of the definition of temporal motif introduced above. 
Illustrated in Fig.~\ref{fig:walkthrough_example} are an input temporal graph $G$, a fine-grained $\delta$-temporal motif $M$ with an anti-edge attached to edge 2, and examples of a valid match and invalid matches.
All timestamps are in units of seconds.
Three potential matches (a), (b), (c) presented in Fig.~\ref{fig:walkthrough_example} are considered invalid because:
(a) the edge $(3, 0, 120)$ happens 110 seconds after edge $(0, 1, 10)$, which is larger than maximum allowed time window $\delta$,
(b) the edge $(0, 1, 150)$ occurs 30 seconds after edge $(3, 0, 120)$, which violates the fine-grained temporal constraint $\delta_3=20$, and
(c) the  edge $(0, 2, 130)$ happens 10 seconds after $(3, 0, 120)$, which is prohibited by the anti-edge $3$ that is attached to edge $2$.
The valid match, on the other hand, satisfies all the requirements specified by the fine-grained $\delta$-temporal motif.

\subsection{Algorithmic Prior Work} \label{subsec:prior-work}
This paper builds upon a state-of-the-art exact mining algorithm proposed by Mackey \textit{et al.}~\cite{mackey2018chronological}, which is more general and efficient compared to other prior works~\cite{paranjape2017motifs}.
In particular, high-speed algorithms presented in Paranjape \textit{et al.}~\cite{paranjape2017motifs} are only applicable to 3-edge star and triangle mining, and their generic algorithm requires mining static motifs first that leads to orders of magnitude more unnecessary work (see Table~\ref{table:vs-static} in \S\ref{sec:results}).
Mackey \textit{et al.}'s algorithm addresses this inefficiency by resolving both structural and temporal constraints at each step of the search process.

Algorithm~\ref{alg:generic_algorithm} presents the pseudocode for Mackey \textit{et al.}'s algorithm.
This algorithm can mine $\delta$-temporal motifs without fine-grained temporal constraints and anti-edges.
The input temporal graph is stored as a chronologically sorted \textit{temporal edge list} as well as data structures that describe its topology in a format similar to the Compressed Sparse Row (CSR).
Two CSR-like data structures are used to maintain incoming and outgoing edges for each vertex, storing their indices in the temporal edge list.
The execution context of the algorithm comprises data structures that maintain vertex mapping information ($MapMG[]$, $MapGM[]$, $eCount[]$), along with a stack ($eStack[]$) to store the indices of matched graph edges.

The algorithm mines temporal motifs by expanding search trees in a Depth-First Search (DFS) manner.
The algorithm starts by mapping the first motif edge with all graph edges, and expanding a search tree to match the rest of the edges iteratively (line~\ref{algo:main_loop}). 
In each \textit{iteration}, the algorithm either finds a match or backtracks.
Each edge in the motif corresponds to a level in the search tree.
In the matching process, the edge is selected from a candidate edge list that is generated (line~\ref{algo:gen_candidates}) based on the nodes $u/v$ in the current partial match.
The list is a subset of $u/v$'s outgoing or incoming edges, denoted as $\mathcal{N}_{out}(u)$ and $\mathcal{N}_{in}(v)$ that also obey temporal constraints.
When a valid match is found, the algorithm updates its execution context and expands the search tree.
Otherwise, it backtracks and voids changes applied to the execution context for the current level.
We refer to the necessary actions to maintain execution context status during search tree expansion and backtracking as \textit{book-keeping} actions.
The indices of the matched edges are maintained as a stack (\textit{eStack[]}).
Mapping information in \textit{MapMG[]} and \textit{MapGM[]} is used to check structural constraints (line~\ref{algo:constraints_checking}) for candidate edges. 

\begin{algorithm}[t]
\scriptsize
\caption{Pseudocode of temporal motif mining algorithm
} \label{alg:generic_algorithm}
\begin{algorithmic}[1]
\State \textbf{Execution Context and Initialization} 
\State \hspace{\algorithmicindent} $MapMG[u] = -1 \; \forall u \in V_M; \; MapGM[u] = -1 \; \forall u \in V_G$
\State \hspace{\algorithmicindent} $eCount[u] = 0 \; \forall u \in V_G,  eStack = [], e_M = -1, e_G = -1, t' \leftarrow \infty$

\State

\Procedure{\textcolor{blue}{TemporalMotifMining}}{$G,\ M,\ \delta$}\\
\hspace*{\algorithmicindent} \textbf{Input}: temporal graph ($V_G, E_G$), motif ($V_M, E_M$), time limit $\delta$.\\
\hspace*{\algorithmicindent} \textbf{Output}: temporal motifs
    \While{true}    \textcolor{blue}{\Comment{Loop until all motifs are found}} \label{algo:main_loop}
        \State $e_G$ = $\textsc{FindNextMatchingEdge}( )$ \textcolor{blue}{\Comment{Search for a valid edge at the current level}}
        \If{$e_G$ is valid} \label{line:valid_edge_found}
            $\textsc{NextLevel}( )$ \textcolor{blue}{\Comment{A valid edge is found and go to the next level}}
        \EndIf
        \State $e_G \; += \; 1$ \textcolor{blue}{\Comment{No more matched edges at the current level}}
        \While{$e_G > |E_G|$ \textbf{or} $time(e_G) > t'$}
        \textcolor{blue}{\Comment{Backtrack until a level with edges to match}}
        \If{$eStack$ is not empty} \label{algo:control11} $\textsc{Backtrack}()$
        \Else $ $ \textbf{return} results 
        \EndIf
        \EndWhile
    \EndWhile
\EndProcedure

\State

\Procedure{\textcolor{blue}{FindNextMatchingEdge}}{ } 
    \State $(u_M, v_M) = E_M[e_M], \ (u_G, v_G) = MapMG[u_M], MapMG[v_M]$
    \State $S \leftarrow \textsc{GetCandidateEdgeList(}u_G, v_G\textsc{)}$\textcolor{blue}{\Comment{Get a list of candidate edges to match}} \label{algo:gen_candidates}
    \For{each edge $e$ in $S$} \textcolor{blue}{\Comment{Return the first edge that satisfies constraints}}
        \If{$\textsc{StructConstraints(}e,u_G, v_G\textsc{)}$ and ${time}(e) < t'$} \label{algo:constraints_checking}
            \State \textbf{return} $e$
        \EndIf
    \EndFor
\EndProcedure

\State

\Procedure{\textcolor{blue}{NextLevel}}{ } \textcolor{blue}{\Comment{Prepare to explore the next level in the search tree}}
    \If{$e_M == |E_M| - 1$}  {\Comment{\textcolor{blue}{Full motif found in the last level}}}
    \State Output a motif $H$ using $eStack$ 
    \Else \textcolor{blue}{\Comment{Partial motif found and expanding the search}}
    \State $\textsc{UpdateDataStructures( )}$; $e_M \; += \;1$ {\Comment{\textcolor{blue}{Bookkeep and push stack}}}
    \State $eStack.push(e_G)$;
    \If{$eStack$ is empty} \textcolor{blue}{\Comment{Set time limit when the first edge is mapped}}
        \State $t' \leftarrow {time}(e_G) + \delta$ 
    \EndIf
    \EndIf
\EndProcedure

\State

\Procedure{\textcolor{blue}{Backtrack}}{ } 
    \State $e_G$ = $eStack.pop()+1$
    \State $\textsc{RollbackDataStructures( )}$; $e_M \; -= \;1$ {\Comment{\textcolor{blue}{Void changes from the context}}}
    \If{$eStack$ is empty}
        $t' \leftarrow \infty$
    \EndIf
\EndProcedure

\State

\Procedure{\textcolor{blue}{StructConstraints}}{$e,\ u_G,\ v_G$} {\Comment{\textcolor{blue}{A valid edge needs to be structurally}}}
    \State $(u_G', v_G') \leftarrow E_G[e]$                      {\Comment{\textcolor{blue}{consistent with the existing partial motif}}}
    \State $u_{consistent} \leftarrow u_G = u_G'$ or $(u_G < 0$ and $MapGM[u_G'] < 0)$ \label{algo:node_consistent_0}
    \State $v_{consistent} \leftarrow v_G = v_G'$ or $(v_G < 0$ and $MapGM[v_G'] < 0)$ \label{algo:node_consistant_1}
    \State \textbf{return} $u_{consistent}$ and $v_{consistent}$

\EndProcedure

\State

\Procedure{\textcolor{blue}{UpdateDataStructures}}{ } \textcolor{blue}{\Comment{Maintain mapping information}}
\State $(u_G, v_G) \leftarrow E_G[e_G], \;\, (u_M, v_M) \leftarrow E_M[e_M]$ 
\State $MapMG[u_M] = u_G, \;\; MapMG[v_M] = v_G$ \textcolor{blue}{\Comment{Map motif node to graph node}}
\State $MapGM[u_G] = u_M, \;\; MapGM[v_G] = v_M$ \textcolor{blue}{\Comment{Map graph node to motif node}}
\State $eCount[u_G] += 1, \;\; eCount[v_G] += 1$ \textcolor{blue}{\Comment{Increment mapped edge cnt}}
\EndProcedure

\State

\Procedure{\textcolor{blue}{RollbackDataStructures}}{ }  \textcolor{blue}{\Comment{Maintain mapping information}}
    \State $eCount[u_G] -= 1, \; eCount[v_G] -= 1   $
    \textcolor{blue}{\Comment{Reduce mapped edge cnt}}
    \If{$eCount[u_G] == 0$}
    \textcolor{blue}{\Comment{No edges of $u_G$ mapped}}
        \State $u_M \leftarrow MapGM[u_G]$
        \State $MapGM[u_G] = -1, \; MapMG[u_M] = -1   $
       \textcolor{blue}{\Comment{Free $u_G,u_M$}}
    \EndIf
    \If{$eCount[v_G] == 0$}
    \textcolor{blue}{\Comment{No edges of $v_G$ mapped}}\label{algo:control12}
        \State $v_M \leftarrow MapGM[v_G]$
        \State $MapGM[v_G] = -1, \; MapMG[v_M] = -1   $
        \textcolor{blue}{\Comment{Free $v_G,v_M$}}
    \EndIf
\EndProcedure

\State

\Procedure{\textcolor{blue}{GetCandidateEdgeList}}{$u_G$, $v_G$} \textcolor{blue}{\Comment{Gather candidates by mapping information}}
    \If{$u_G \geq 0$ and $v_G \geq 0$} \textcolor{blue}{\Comment{Both $u_G, v_G$ mapped to motif nodes}}\label{algo:control21}
        \State $S\ \leftarrow \{ e \in \mathcal{N}_{out}(u_G)/\mathcal{N}_{in}(v_G) : t_e > {time}(e_G) \}$ \textcolor{red}{\Comment{filter via binary search}} \label{algo:bsearch0}
    \ElsIf{$u_G > 0$} \textcolor{blue}{\Comment{Only $u_G$ mapped to a motif node}}
        \State $S\ \leftarrow \{e \in \mathcal{N}_{out}(u_G) : t_e > {time}(e_G) \}$  \textcolor{red}{\Comment{filter via binary search}} \label{algo:bsearch1}
    \ElsIf{$v_G > 0$} \textcolor{blue}{\Comment{Only $v_G$ mapped to a motif node}}
        \State $S\ \leftarrow \{e \in \mathcal{N}_{in}(v_G) : t_e > {time}(e_G) \}$  \textcolor{red}{\Comment{filter via binary search}} \label{algo:bsearch2}
    \Else \textcolor{blue}{\Comment{Both $u_G,v_G$ not mapped}} \label{algo:control22}
        \State $S\ \leftarrow \{e \in E_G: t_e > {time}(e_G)  \}$  
    \EndIf
    \State \textbf{return} $S$

\EndProcedure

\end{algorithmic}
\end{algorithm}



\subsection{GPU Architecture} \label{subsec:gpu_architecture}
To make our optimizations accessible to readers, we briefly discuss the basics of GPU architecture.
GPUs offer significantly higher compute throughput and memory bandwidth compared to CPUs.
A GPU consists of tens of Streaming Multiprocessors (SMs).
Host CPU launches kernels that contain thousands of parallel threads for GPU execution.
GPU \textit{threads} are grouped into \textit{blocks} to execute on the SMs.
Threads in a block are executed in a Single Instruction Multiple Threads (SIMT) execution model.
Threads among the same block are grouped into \textit{warps} to be executed together.
GPU achieves peak throughput when all threads are on the same execution path.
However, when the threads in a warp take a branch, their execution path may diverge.
This is known as \textit{branch divergence}, where a GPU needs to execute the resulting two paths sequentially, which reduces overall performance. 
Branch divergence leaves GPU hardware to be idle when a subset of threads is not executed in each divergent path.
Besides, non-uniform amount of work assigned to GPU threads under-utilizes compute resources.

GPUs are equipped with DRAM as their \textit{device memory} to hold data for execution, the size of which is often limited (in 10s of GB).
DRAM in GPU can be accessed by high-bandwidth interfaces such as GDDRx.
To hide the memory latency of fetching data from DRAM, GPUs also employ on-chip cache memory and large register files.
This physical memory hierarchy is used by several virtual memory spaces, \textit{e.g.,} constant memory and shared memory.
Read-only data structures, \textit{e.g.,} input temporal graph, can be placed in constant memory.
Shared memory, on the other hand, can be used to enable communication between different threads within a block.
Furthermore, threads in the same warp can also communicate via register shuffling at a much lower cost.

\section{Motivation}



\subsection{High Algorithmic Complexity} \label{subsec:complexity}
Exactly mining temporal motifs poses significant computational challenges due to their inherent complexity. Denote by $|E_G|$ the number of temporal edges, and by $|E_M|$ the number of temporal edges and anti-edges in the motif, respectively. 
Let $k$ be the expected number of edges occurring in the graph within a duration $\delta$. 
The algorithmic complexity of generalized temporal motif mining is $\mathcal{O}\left( |E_G| \cdot k^{|E_M| - 1}\right)$: it grows \textit{exponentially} with the size of the motif, \textit{polynomially} with the average number of edges occurring within time $\delta$, and \textit{linearly} with the size of the temporal graph. 
Extending the time constraint $\delta$ increases the average number of edges occurring within the time duration. This means that for each edge in $E_M$, there will be more edges to explore in $E_G$, thereby extending the \textit{width} of the DFS search tree. However, adding fine-grained temporal constraints \textit{reduces} the width of the search tree. 
Increasing the size of the motif increases the \textit{depth} of the search tree.
In the worst case, there are $|E_G|$ edges to match for each edge added to the motif, thus the complexity grows exponentially with the tree depth. 
Notably, this also holds for adding anti-edges because in the worst case, one needs to explore $|E_G|$ edges to ensure the exclusion of the anti-edge.
%

\subsection{Suboptimal Performance of Existing Software on GPUs} \label{subsec:GPU_baseline_bottlenecks}
GPU is an attractive candidate to accelerate the complex temporal motif mining workload due to its high compute throughput and memory bandwidth.
Prior works~\cite{pangolin,G2Miner2022OSDI} have already made a strong case for accelerating static graph mining on GPUs.
However, no work exists that optimizes motif mining in temporal graphs.
Next, we study the performance of Mackey \textit{et al.}'s algorithm on a GPU.

\textbf{Baseline GPU Implementation.}
It is trivial to parallelize Algorithm~\ref{alg:generic_algorithm} by expanding each search tree using a different thread.
However, the memory complexity of \textit{MapGM[]} and \textit{eCount[]} is $O(|V_G|)$, making it infeasible to allocate this large amount of memory for each thread on a GPU.
Our GPU baseline implementation removes \textit{MapGM[]} and only uses \textit{MapMG[]} to check the structural constraints (line \ref{algo:constraints_checking}).
The entire \textit{MapMG[]} table is iterated to verify whether a new edge is consistent with the current partial result, instead of using \textit{MapGM[]} (line~\ref{algo:node_consistent_0} - \ref{algo:node_consistant_1}).
On the other hand, \textit{eCount[]} can only use $|V_M|$ entries by slightly changing the book-keeping logic in \textsc{UpdateDataStructures()} and \textsc{RollbackDataStructure()} subroutines.
The resulting GPU implementation expands an independent search tree on each thread.

\begin{figure}[t]
\centerline{\includegraphics[width=\linewidth]{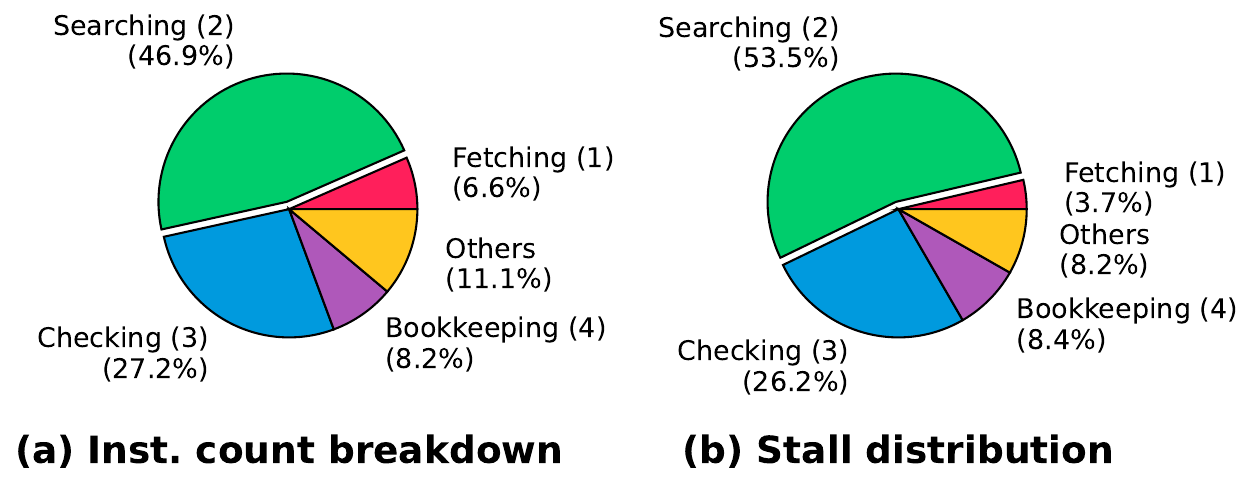}}
\vspace{-0.35cm}
\caption{Profiling results for mining \texttt{M6} on \WIKI: (a) instruction count breakdown and (b) stall distribution.}
\vspace{-0.65cm}
\label{fig:motivation_prof}
\end{figure}

\textbf{Long Execution Times.}
Using this implementation, we measure the time to find temporal motifs using an NVIDIA A40 GPU.
We refer the reader to \S\ref{sec:methodology} for details on input temporal motif queries and temporal graphs.
Using our experimental setup, we find that mining tasks run for several hours (\textit{e.g.,} mining \texttt{M6}, \texttt{M11}, and \texttt{M12} in \ETH\ graph takes 18.6 hours, 3.3 hours, and 3.4 hours, respectively).
While we use some of the largest graph datasets available in the public domain, larger graphs used by the industry further exacerbate runtime, motivating our optimization effort.



\textbf{Execution Bottlenecks.}
Using our baseline implementation, we conduct a detailed performance analysis to find execution bottlenecks as detailed below.
We use a representative workload of mining \texttt{M6} in \WIKI\ graph to conduct this study.

\textbf{\textit{1. Long-latency binary search operations.}} \label{subsubsec:bsearch}
In order to generate a list of candidate edges, Algorithm~\ref{alg:generic_algorithm} uses binary search to filter out neighboring edges of a vertex that does not satisfy temporal constraints (lines \ref{algo:bsearch0}, \ref{algo:bsearch1}, \ref{algo:bsearch2}).
This induces random accesses to the input graph's large CSR structures which are not GPU cache friendly.
This results in long latency device memory accesses that stall GPU execution.
Fig.~\ref{fig:motivation_prof} presents quantitative evidence for this.
In this figure, fetching and searching correspond to fetching neighboring edges and performing binary search, respectively, in the \textsc{GetCandidateEdgeList()} function.
Fig.~\ref{fig:motivation_prof}(a,b) show that binary search contributes to 53.5\% of GPU instructions and 57.2\% of stall cycles.
Furthermore, these long-latency memory accesses contribute to 56.1\% stalls during the search phase (not shown due to space limit).

\begin{figure}[t]
  \centering
  \includegraphics[width=\linewidth]{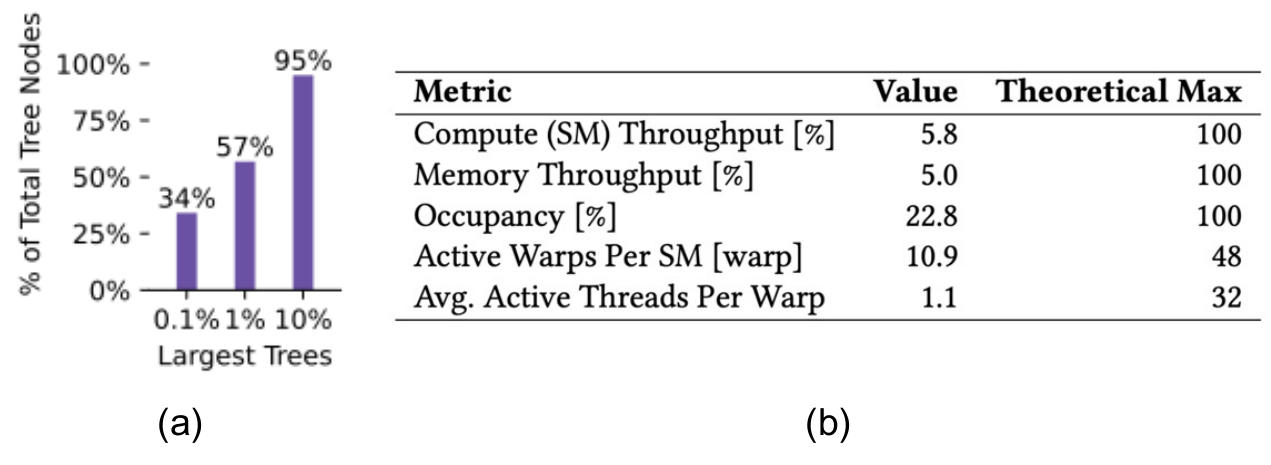}
  \vspace{-0.7cm}
  \caption{(a) Non-uniform size of search trees: percentage of nodes in the largest search trees among all search tree nodes, (b) effect of load imbalance on GPU performance metrics.}
  \label{fig:work_distribution}
\end{figure}


\textbf{\textit{2. Excessive thread divergence.}}
As explained in \S\ref{subsec:gpu_architecture}, GPU handles branch/if conditions in the code using the branch divergence mechanism that leaves a subset of compute units mapped to a warp idle.
Algorithm~\ref{alg:generic_algorithm} clearly shows that branch operations are ubiquitous in the temporal motif mining algorithm.
Furthermore, the outcome of these branch operations changes as the algorithm progresses as they depend on dynamic runtime data.
For example, the \textsc{GetCandidateEdgeList()} function uses the information of which graph nodes are previously mapped to motif nodes to decide which neighborhood information to fetch from memory.
The widespread nature of these branch operations results in frequent branch divergence, and consequently, under-utilization of GPU resources.

\textbf{\textit{3. Extreme load imbalance.}}
The baseline implementation expands each search tree using a unique GPU thread.
The amount of work for expanding the search tree depends on the motif and graph structures, and the extent of temporal interactions between nodes.
Because real-world graphs follow power-law connectivity structure (\textit{i.e.,} a small subset of popular nodes connected to a large fraction of edges) the size of each search tree, consequently, the amount of work assigned to each thread varies significantly.
Fig.~\ref{fig:work_distribution}(a) shows that the top 0.1\% of the search trees (by size) constitute 34\% of explored tree nodes in the entire execution.
This clearly shows that the work assignment to each thread is \textit{highly skewed}.
Fig.~\ref{fig:work_distribution}(b) shows the effect of this on GPU performance metrics.
Because of work imbalance, this implementation only achieves 22.8\% occupancy, and has 10.9 active warps per SM, much lower than theoretical limits.
Furthermore, on average, only 1 out of 32 threads are active.

Next, we briefly explain the types of this load imbalance at different levels of the GPU architecture.
All threads within a GPU warp use same instructions to execute in a SIMT fashion; when certain threads complete mining their assigned search trees before others, this leads to \textit{intra-warp load imbalance}.
All threads within a GPU block are implicitly synchronized; the compute resources allocated to a block cannot be released until all threads return.
When a subset of threads explore large search trees while others finish execution, this also results in \textit{intra-block load imbalance}, rendering other GPU resources idle.
A GPU kernel finishes when all the blocks finish execution.
With extreme load imbalance, there are often a few blocks with outstanding work, while others finish, that render GPU resources idle, causing \textit{intra-kernel-launch load imbalance}.

\subsection{Need For More Expressive Queries}
It is possible to make the definition of $\delta$-temporal motifs more expressive using additional constraints.
Certain algorithms~\cite{kovanen2011temporal,pasha21Kdd} for mining temporal motifs adopt more detailed temporal constraints on consecutive edges within the motifs.
In some static pattern mining systems like \cite{peregrine}, the concept of anti-edge is introduced to let users further restrict their patterns.
Missing these constraints from users leads to unwanted results that increase execution time for searching and post-processing.
An ideal system should support the most expressive query definition that includes labels on nodes/edges, fine-grained temporal constraints, and temporal anti-edge.
This abundant set of constraints empowers users to accurately express their intention for practical real-world mining use cases.

\section{\THISWORK\ Design Overview} \label{sec:design_overview}




This section provides the design goals and overview of the execution pipeline, user-defined query, and runtime components of \THISWORK.

\subsection{Design Goals}
\label{subsec:design_goals}



\THISWORK\ addresses the unique challenges of efficiently executing temporal motif mining on multi-GPU systems with the following specific design goals.

\textbf{GPU-friendly algorithm and implementation.}
Naively mapping Mackey \textit{et al.}'s algorithm to GPU causes sub-optimal execution behavior (\S\ref{subsec:GPU_baseline_bottlenecks}).
The presence of long-latency binary search and control flow operations significantly declines the overall performance.
Therefore, \THISWORK\ needs to integrate GPU-friendly adaptations into the original algorithm and provide high-quality implementation to enhance high efficiency on GPUs.

\textbf{Memory footprint reduction and multi-GPU support.}
While a single GPU offers massive throughput, the amount of memory on each GPU is often limited.
For example, a state-of-the-art GPU used for our evaluation, \textit{i.e.,} NVIDIA A40, only has 48GB of memory.
Therefore, it is imperative to reduce the amount of memory footprint used by the workload.
Furthermore, many real-world graphs are extremely large in size, entailing billions to trillions of nodes and edges.
This precludes a graph input itself to fit on a single GPU's memory.
Therefore, important design goals of \THISWORK\ include reducing the memory footprint of the workload and offering multi-GPU support to scale to large input graph sizes.

\textbf{Load balancing at all levels.}
The skewed work distribution of temporal motif mining results in harmful load imbalance at every level of GPU execution (\S \ref{subsec:GPU_baseline_bottlenecks}).
Because the degree of imbalance depends on both the input temporal graph and user-defined query, the imbalance can lead to exceedingly poor system performance in certain scenarios.
A key design goal of \THISWORK\ runtime system is to be able to intelligently detect load imbalance and execute load re-distribution at runtime at all levels of the GPU execution to ensure consistently high overall performance.

\textbf{Expressive and user-friendly query interface.}
While a few prior works~\cite{paranjape2017motifs,mackey2018chronological,kumar20182scent,wang2020efficient,liu2018sampling,sarpe2021presto} optimize the execution of temporal motif mining on CPUs, they can only support limited motif query definitions (\textit{i.e.,} $\delta$-temporal motif discussed in \S\ref{subsec:problem_def}).
\THISWORK\ strives to adopt a broader definition of temporal motif, supporting additional features such as fine-grained temporal constraints, labels on both nodes and edges, and temporal anti-edges.
Additionally, \THISWORK\ supports both motif counting and enumeration.
Such a comprehensive and user-friendly interface is essential to enhance the system's usability.

\subsection{Execution Pipeline} \label{subsec:execution_pipeline}
\begin{figure}[t]
    \centering
    \includegraphics[width=\linewidth]{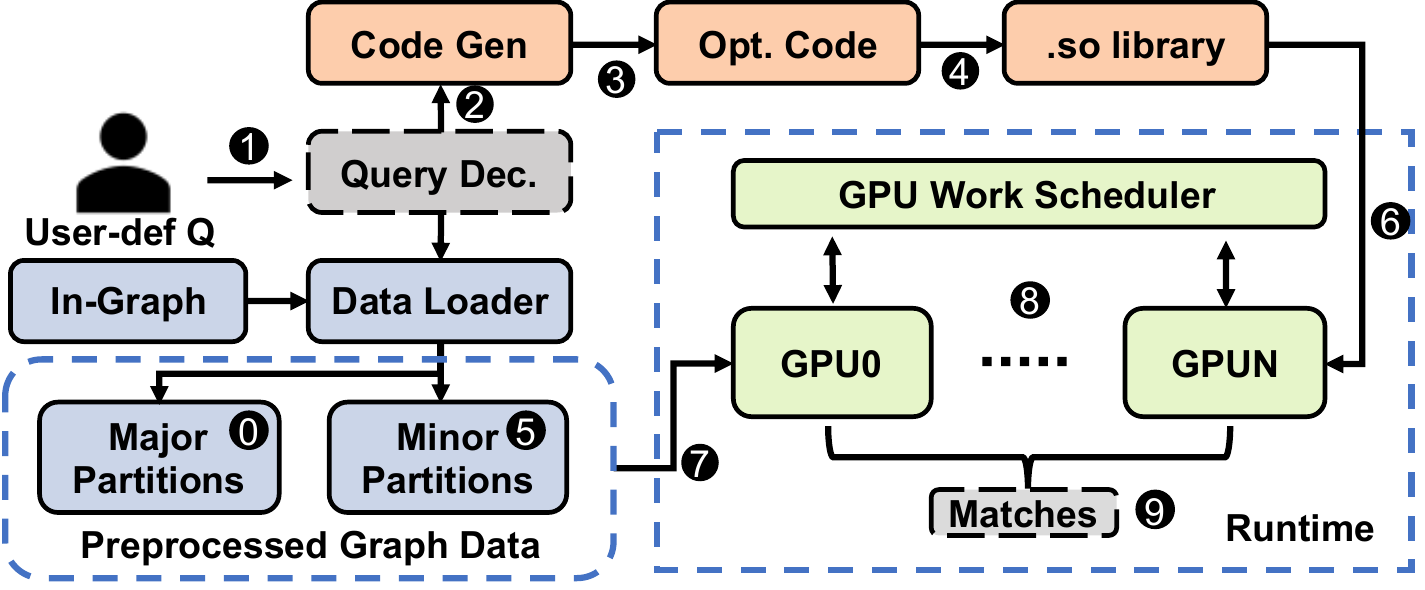}
    \vspace{-0.7cm}
    \caption{\THISWORK\ execution pipeline overview. Input: temporal graph and user-defined query. Output: optimized runtime and pre-processed data graph for multi-GPU execution.}
    \vspace{-0.45cm}
    \label{fig:design-high-level}
\end{figure}
Fig.~\ref{fig:design-high-level} provides a high-level overview of the execution pipeline of \THISWORK.
The inputs to this pipeline include a temporal graph and a user-defined query.
The outputs of this pipeline are an optimized execution plan and intelligent runtime, and pre-processed graph, which are used to mine user-defined queries within an input graph on a multi-GPU system.
Single-GPU system is a special case of the multi-GPU system, and can be defined by the user.

Before a query is decoded, an input graph is processed. 
Following the partition strategy elaborated later in \S\ref{subsubsec:edge_list_partitioning}, the data loader generates \textit{major partitions} from the temporal edge list (\circled{0}).
The generation of major partitions occurs prior to any query processing and is not on the critical path of resolving queries.
The data loader uses GPUs to perform all the data preprocessing for high efficiency.
Next, the user-defined query (\circled{1}) is decoded and inputted to the code generator (\circled{2}).
Note that \THISWORK\ supports the most generalized definition of a temporal motif with an arbitrary number of nodes/edges/connections, fine-grained temporal constraints, node/edge labels, and anti-edges.
The output of mining can be either motif enumeration or counting.
The code generator generates optimized code that uses the minimum  number of runtime variables for the mining to improve occupancy. 
This also includes multiple runtime mechanisms to improve the execution efficiency and routines to handle additional constraints (\circled{3}), as  detailed in \S\ref{subsec:runtime_overview}.
The optimized code is compiled into a shared library (.so) (\circled{4}) that will be later loaded into \THISWORK\ runtime.
At the same time, \textit{minor graph partitions} are constructed based on the partitioning strategy (\circled{5}, \S\ref{subsubsec:edge_list_partitioning}) for a given query.

The runtime system loads the optimized library (\circled{6}), and uses it to construct \textit{GPU workers} on each GPU.
Instructed by the \textit{GPU work scheduler}, each GPU worker loads and processes a different graph partition (\circled{7}).
The GPU work scheduler assigns work to GPU workers, monitors their execution status, and performs load balancing if necessary (\circled{8}, \S\ref{subsubsec:mgpu-sched}).
After all the partitions are consumed, the results are collected from all GPU workers (\circled{9}).
The results include motif matches (enumeration or counting) for a given user-defined query on an input temporal graph.

\subsection{User-Defined Query} \label{subsec:query_overview}
\begin{figure}[t]
    \centering
    \includegraphics[width=\linewidth]{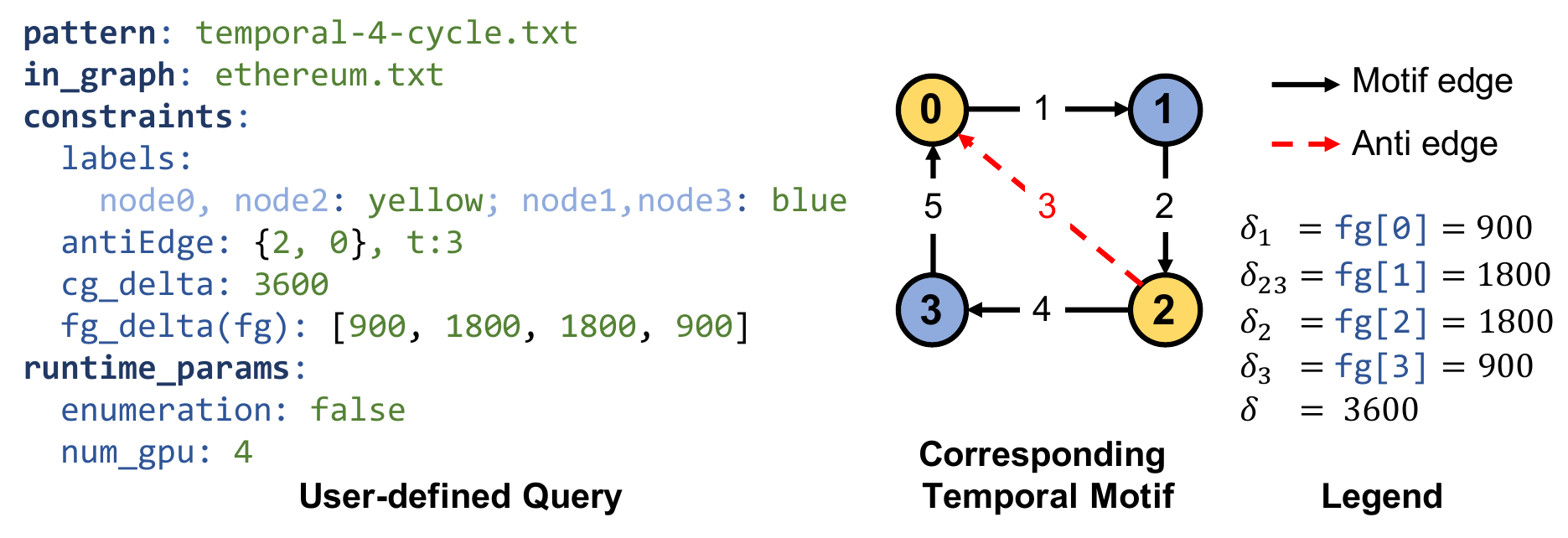}
    \vspace{-0.75cm}
    \caption{Example user-defined query to mine a temporal 4-cycle with fine-grained temporal constraints and an anti-edge in an input graph \texttt{ethereum}.}
    \label{fig:query_example}
\end{figure}
Fig.~\ref{fig:query_example} shows the user-defined query input.
There are four sections in this query: 1) temporal motif pattern (\texttt{pattern}), 2) input temporal graph (\texttt{in\_graph}), 3) additional constraints (\texttt{constraints}), and 4) runtime parameters (\texttt{runtime\_params}).
The first section details the basic shape and connectivity of a temporal motif to be mined.
While an actual query details the structure and temporal order of motif edges, for brevity, we show it as a temporal 4-cycle.
Additionally, the query specified the file location of an input temporal graph.

The \texttt{constraints} section details additional constraints to be mined.
Following the formal definitions of these constraints from \S\ref{subsec:problem_def}, this example query attempts to mine a motif with the following labels: two yellow and two blue nodes.
While these labels are simplified as node colors for illustration, it is possible to specify more realistic labels in \THISWORK\ (\textit{e.g.,} customer/merchant labels for financial transaction network) for real-world graph settings.
Additionally, a temporal anti-edge is specified with a temporal order (\texttt{t:3}).
This means an anti-edge is attached to edge \texttt{2} and the motif matches must not contain an edge between nodes \texttt{2} and \texttt{0} within a time window ($\delta_{23}$) from edge \texttt{2}.
This follows temporal constraints: 1) course-grained (\texttt{cg\_delta}) and 2) fine-grained (\texttt{fg\_delta}).
The course-grained temporal constraint is the time window ($\delta$) for an entire motif.
Fine-grained constraints, on the other hand, denote maximum time difference between consecutive edges (\textit{e.g.,} $\delta_1$).

The last section specifies two runtime constraints.
The example in Fig.~\ref{fig:query_example} shows that a user is interested in counting the number of motifs (\textit{i.e.,} enumeration is false).
In the case that enumeration is true, a user must define the number of matches to be enumerated (not shown in the figure).
Furthermore, a user can also define the number of GPUs (default is set to one).


\subsection{Runtime Components} \label{subsec:runtime_overview}

\begin{figure}[t]
    \centering
    \includegraphics[width=\linewidth]{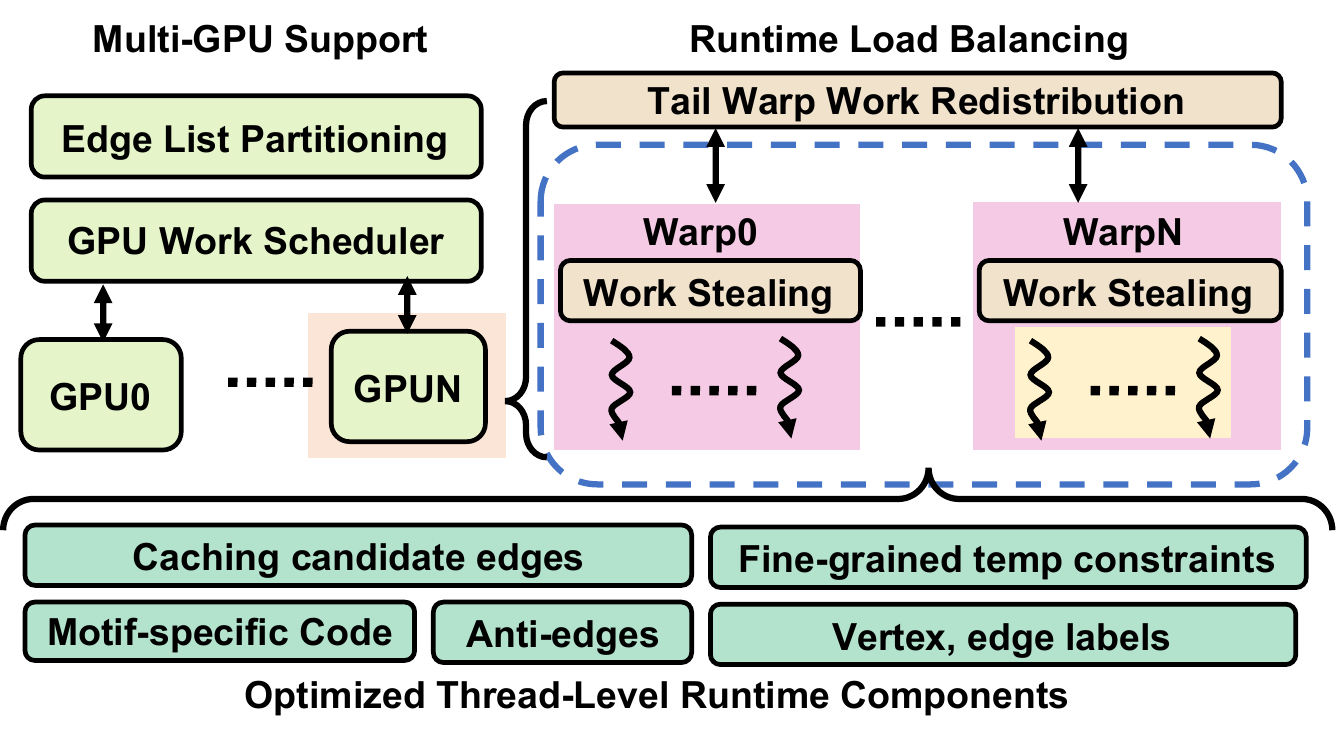}
    \vspace{-0.8cm}
    \caption{\THISWORK\ runtime components overview.}
    \label{fig:runtime-high-level}
\end{figure}
Fig.~\ref{fig:runtime-high-level} presents the runtime components as outlined below.

\subsubsection{Thread-local Components.} \label{subsubsec:thread_local_comp_outline}
Each GPU thread explores a search tree to mine a user-defined motif in an input graph.
\THISWORK\ GPU code supports two types of runtime components that execute at thread level as shown in green boxes in Fig.~\ref{fig:runtime-high-level}: 1) unique structural and temporal constraints from the query, and 2) runtime optimizations.
As discussed in \S\ref{subsec:query_overview}, motif constraints include fine-grained temporal constraints, temporal anti-edges, and labels on vertices and/or edges.
Additionally, \THISWORK\ proposes the following two optimizations to improve performance.

\noindent \textbf{Caching candidate edges (\S\ref{subsubsec:candidate_edges}).} 
To avoid unnecessary binary search operations during backtracking, per-thread context caches the candidate edges at each search level. 
Moreover, the cached candidate edge entries enable work sharing among threads.

\noindent \textbf{Motif-specific code generation (\S\ref{subsubsec:decoding}).}
While our system supports mining any arbitrary motifs, the proposed \THISWORK\ backend generates motif-specific optimized code for GPU execution.
In particular, \THISWORK\ generates code with auxiliary data structures to simplify the program's control flow at runtime.

\subsubsection{Load balancing at a Single-GPU Level.}
Beyond the scope of a single GPU thread, the following components enable the detection and re-distribution of skewed workload at a single-GPU level to improve hardware utilization.


\noindent \textbf{Intra-warp work stealing (\S\ref{subsubsec:intra_warp_sharing}).}
This mechanism takes advantage of intra-warp fast register shuffling and synchronization primitives to perform frequent work stealing at a low cost.


\noindent \textbf{Tail warp work redistribution (\S\ref{subsubsec:tail_expan}).}
This optimization redistributes the work among all warps when there is no pending work to keep all warps busy within a kernel launch.
It aborts warps that are likely exploring large search trees and share their work with other idle warps.

\subsubsection{Multi-GPU support.}
\THISWORK\ provides a low-cost edge list partition strategy to divide work for multiple GPUs.
At runtime, a GPU work scheduler employs a queue-based work scheduling to balance the load among multiple GPUs.




\section{\THISWORK\ Design and Optimizations} \label{sec:design_details_optimizations}

This section discusses \THISWORK\ design and optimizations in detail.
\THISWORK\ completely hides the complexity of the proposed optimizations from a user.
As discussed in \S\ref{subsec:execution_pipeline}, these optimizations are conducted as part of \THISWORK\ backend.

\subsection{Execution Efficiency Optimizations}
Below, we discuss thread-local runtime optimizations.
\begin{figure}[t]
  \centering
  \includegraphics[width=\linewidth]{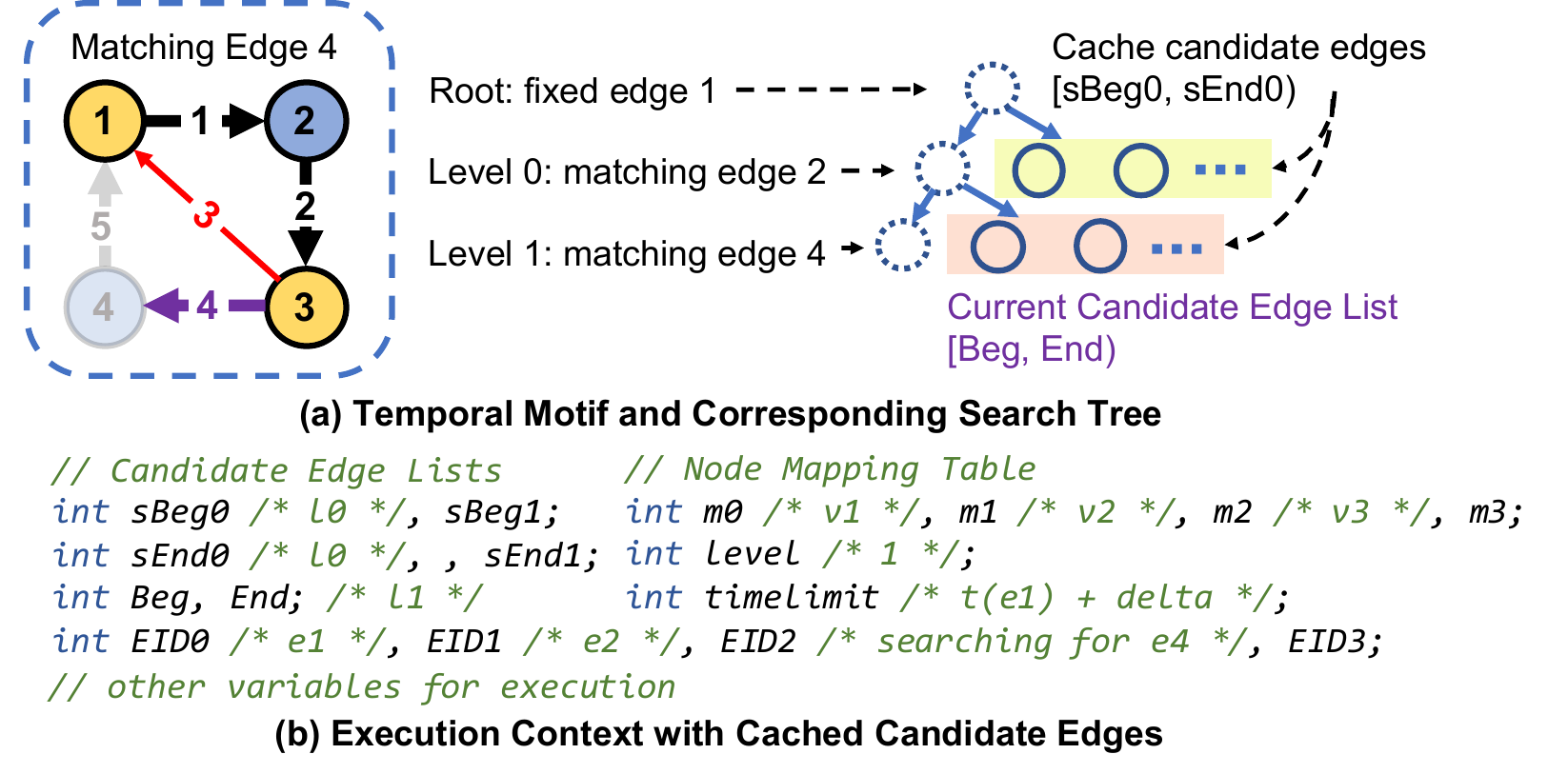}
  \vspace{-0.75cm}
  \caption{(a) Input temporal motif and its corresponding search tree expansion with the proposed cached candidate edges, and (b) per-thread execution context while matching motif edge 4.}
  \vspace{-0.5cm}
  \label{fig:first-two}
\end{figure}
\subsubsection{Caching Candidate Edges.} \label{subsubsec:candidate_edges}
In the baseline algorithm, when a thread expands the search tree to the next level, it discards the candidate edges at the current level (line~\ref{line:valid_edge_found} in Algorithm~\ref{alg:generic_algorithm}).
As a result, when the algorithm backtracks, it needs to search for these edges again by performing expensive binary search that constitutes around 50\% of the total instructions and stalls (\S\ref{subsec:GPU_baseline_bottlenecks}).


\textbf{Proposed Optimization.}
Based on this observation, we find an opportunity to reduce the number of costly binary search operations in the workload by caching candidate edges.
Specifically, when a match between a motif edge and a graph edge is found, we propose to cache the rest of the candidate edges that could have been matched in a thread context.
The result of this caching is used while backtracking from the same tree node.
This converts a costly binary search operation into a thread context read while backtracking.
This results in reducing the number of binary search operations by half at a marginal cost of thread context storage increase.

Fig.~\ref{fig:first-two}(a) illustrates our solution, which shows the search tree status when the algorithm is matching edge 4 in the motif.
The search tree has two levels.
Level 0 corresponds to the motif edge 2 and level 1 is searching for an edge in its candidate edge list to match the motif edge 4.
The candidate edge list at the current search level (level 1) is represented by \texttt{Beg} and \texttt{End}.
Before the search tree expands to level 1, the remaining candidate edges in level 0 are stored in \texttt{sBeg0} and \texttt{sEnd0}.
When the candidate edge list in level 1 is exhausted, the algorithm backtracks and loads the remaining candidate edge list to \texttt{Beg} and \texttt{End} to continue execution at level 0, avoiding a costly edge search operation.

\subsubsection{Motif-Specific Code Generation.} \label{subsubsec:decoding}
The baseline generalized algorithm supports mining of any arbitrary $\delta$-temporal motif, precluding specific motif-specific optimizations.
While \THISWORK\ is a versatile mining system supporting any arbitrary temporal motif, the proposed code generator is further specialized to employ additional optimizations for user-specific queries.

\textbf{Proposed Optimizations.}
After decoding a specific input motif, \THISWORK\ generates a read-only data structure called \texttt{minfo} before the execution starts.
This is an array of structures whose an $i^{th}$ entry describes the execution progress of the $i^{th}$ level of the search tree.
In what follows, we illustrate how to use this data structure to simplify various mining operations including book-keeping, candidate edge generation, and structural constraint checking.
As discussed in \S\ref{subsec:GPU_baseline_bottlenecks}, these operations result in excessive thread divergences and under-utilize GPU resources.

\textbf{1. Simplifying Book-keeping Operations.}
Using \texttt{minfo}, it is possible to specify the number of valid mappings at each search tree level.
Using this, we can remove the costly operations to vertex mapping data structures.
For example, Fig.~\ref{fig:first-two}(b) shows that \texttt{m0-m3} store node mapping information.
Among them, only the first three are valid at level 1.
When edge 4 is being searched, only vertices 1, 2, and 3 are mapped for the partial motif.
At each book-keeping phase, \texttt{minfo} is used to avoid unnecessary operations to maintain vertex mapping data structures by keeping the number of valid mappings at each level.
Voiding changes to mapping structures during backtracking is automatic by loading the \texttt{minfo} entry at the lower level.
Using this optimization, \THISWORK\ completely removes \texttt{eCount[]} and sub-routine \textsc{RollbackDataStructures} from Algorithm~\ref{alg:generic_algorithm}.


\textbf{2. Simplifying Edge List Generation.}
In Algorithm~\ref{alg:generic_algorithm} the candidate edge lists are generated in \textsc{GetCandidateEdgeList} based on the runtime information of which graph vertices are previously mapped.
As vertex mappings depend on the progression of the algorithm, this branching behavior results in thread divergence and performance slowdown.
\THISWORK\ proposes to use \texttt{minfo} to statically record where to get candidate edges for each search tree level based on the structure and temporal order of motif edges.
This way, the costly branching behavior is transformed into \texttt{minfo} reads in most cases.
The only exception to this is when both source and destination vertices are mapped, where \THISWORK\ performs vertex mapping-based edge list selection.

\textbf{3. Simplifying Structural Constraint Checking.}
In the GPU baseline, the whole mapping table is iterated over to check structural constraints for a new edge.
This can be greatly simplified with \texttt{minfo}.
For example, to match edge 4 in Fig.~\ref{fig:first-two}, the algorithm only needs to check whether its destination is conflicting with the mapping for vertex 1 and 2 (\texttt{m0-m1}).
\texttt{minfo} simplifies structural constraint checking by using pre-decoded motif structure and temporal edge order.
Thus, fewer than half of the comparisons are used to check structural constraints. 

\subsubsection{Lifting Thread Context to Registers.}

We lift thread contexts to store them into the GPU registers to enable faster access and the intra-warp work stealing mechanism (\S\ref{subsubsec:intra_warp_sharing}).
The array-like data structures including the stacks and the mapping table are declared as a set of variables and be accessed via \texttt{switch} statements, such as \texttt{sBeg}, \texttt{sEnd}, and \texttt{m} variables.

\subsection{Load Balancing Optimizations} \label{subsec:load_balancing_opt}
The baseline implementation maps the expansion of each search tree to a GPU thread.
This coarse-grained mapping exacerbates load imbalance as different search trees explore vastly different numbers of tree nodes (Fig.~\ref{fig:work_distribution}).
Next, we introduce the concept of Sub-Tree-Level Parallelism (STLP) and load balancing in \THISWORK.

\textbf{Sub-Tree-Level Parallelism (STLP).}
We refer to expanding sub-trees of a search tree in parallel as \textit{sub-tree-level parallelism}.
Due to the DFS search order, it is not possible to exploit STLP in Mackey \textit{et al.}~\cite{mackey2018chronological}'s algorithm as the nature and sizes of sub-trees within the same tree are unknown at runtime.
However, a sub-tree at each level can be expanded from the candidate edges (\S\ref{subsubsec:candidate_edges}).
\THISWORK\ proposes to \textit{speculatively} exploit STLP by sharing edges inside the candidate edge list with multiple threads.
Specifically, \THISWORK\ speculates that the shared candidate edges imply unexplored sub-trees.
The speculative nature of this optimization may result in unnecessary work when the candidate edges fail to meet structural/temporal constraints, and thus discarded in the original algorithm.
Below, we present two sophisticated optimizations and their implementations that intelligently employ STLP without hurting overall performance.

\subsubsection{Intra-warp Work Stealing.}\label{subsubsec:intra_warp_sharing}
\begin{figure}[t]
  \centering
  \includegraphics[width=\linewidth]{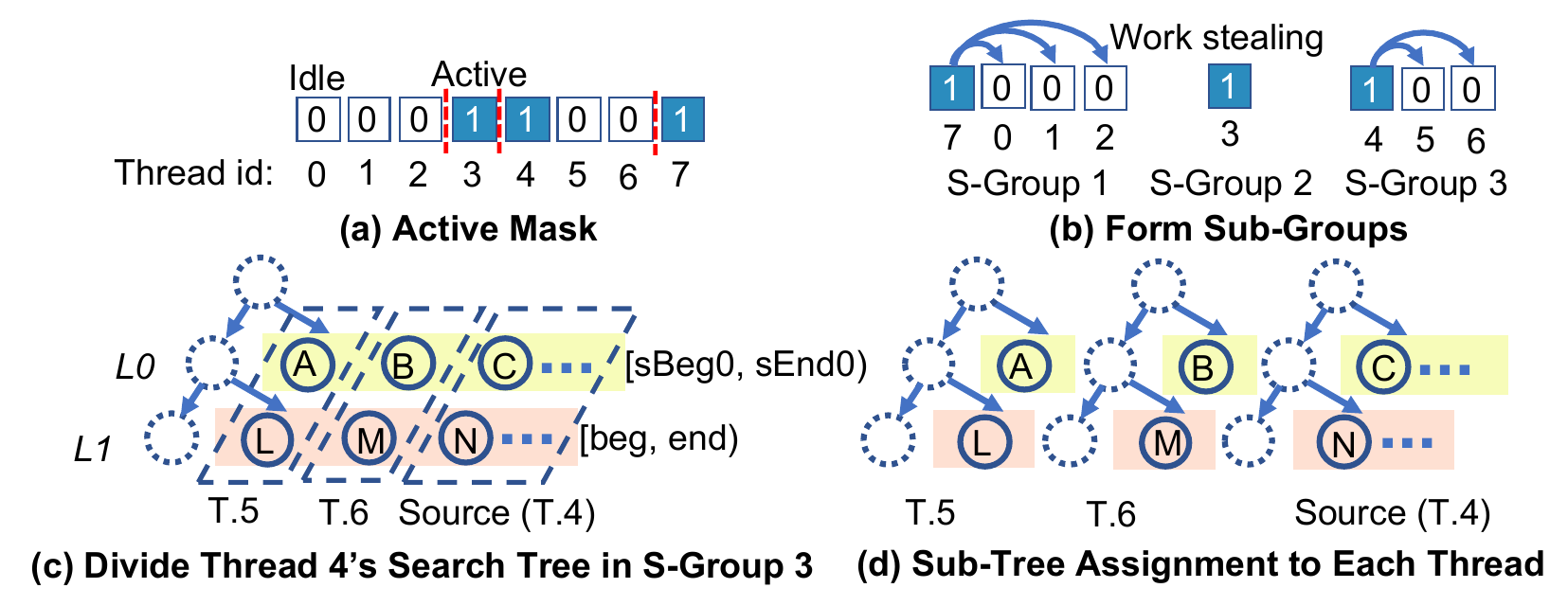}
  \vspace{-0.8cm}
  \caption{Example of intra-warp work stealing.}
  \vspace{-0.5cm}
  \label{fig:intra-warp}
\end{figure}

The goal of intra-warp work stealing is to reduce the impact of intra-warp work imbalance using sub-tree-level parallelism.
On average, the GPU baseline only keeps 1 thread active in each warp due to the imbalanced load (\S\ref{subsec:GPU_baseline_bottlenecks}).
\THISWORK\ proposes a novel low-cost work stealing mechanism using warp-level primitives in CUDA.
The proposed design introduces minimum overheads and selects candidate edges that are less likely to be discarded due to constraint violations.



We define a thread to be \textit{active} or \textit{idle} depending on whether it has outstanding work or not.
At the end of each iteration, the presence of any idle threads triggers this work stealing mechanism.
Empirically, we enable this after 20 iterations to avoid slowing down warps that have no threads with much outstanding work.
To realize this mechanism, all threads use \texttt{\_\_balloc\_sync} to generate an \textit{Active Mask}, where $i^{th}$ bit denotes whether the $i^{th}$ thread is active.
Using this mask, all threads in a warp are divided into multiple subgroups, each with one active thread.
Using this thread grouping, an active thread shares work to other idle threads within the same subgroup, as explained with an example below.

Fig.~\ref{fig:intra-warp}(a,b) illustrates how threads are combined in three subgroups, where each subgroup has one active thread.
Within each subgroup, an idle thread will steal one candidate edge from the beginning of the candidate edge list at each level.
It also copies other execution contexts from the active thread.
The stealing is conducted via register shuffling to  minimize overhead.
The edges are shared from the beginning of edge lists such that they are the most likely to satisfy the temporal constraint.
Fig.~\ref{fig:intra-warp}(c,d) further illustrates this work stealing with an example of subgroup 3.
The first two edges in the candidate edge list at level 0 are \texttt{A} and \texttt{B}, and they are stolen by threads 5 and 6, respectively.
Similarly, thread 5 and 6 steal edge \texttt{L} and \texttt{M} from level 1.
After this work stealing, threads 5 and 6 independently expand their sub-trees in parallel.

\textbf{Theoretical Performance Analysis.}
The warp-level speedup of this optimization can be roughly modeled as 
$
\frac{32}{ t_\text{imb} \left( 1 + \frac{k\epsilon}{\mathcal{I}_\text{opt}} \right) }
$, where $\mathcal{I}_\text{opt}$ is the number of search iterations used by the optimized code within a warp.
$t_\text{imb} < 32$ is the number of active threads within a warp in the imbalanced baseline.
$k < \mathcal{I}_\text{opt}$ is the number of times this optimization is triggered and $\epsilon < 1$ is the work-stealing overhead relative to a search iteration. 
We observe that $t_{\text{imb}}$ is usually much smaller than warp width 32 due to intra-warp work imbalance.
Each trigger of this optimization steals multiple sub-trees, resulting in a smaller $k$.
Our warp-primitive-based efficient implementation results in a smaller $\epsilon$.
In practice, we observe $t_\text{imb} \left( 1 + \frac{k\epsilon}{\mathcal{I}_\text{opt}} \right) < 32$, and a meaningful speedup on average across warps.
Furthermore, the speedup is correlated to the work, $\mathcal{I}_\text{opt}$, assigned to a warp.

\subsubsection{Tail Warp Work Redistribution.}
\label{subsubsec:tail_expan}

\begin{figure}[t]
  \centering
  \includegraphics[width=\linewidth]{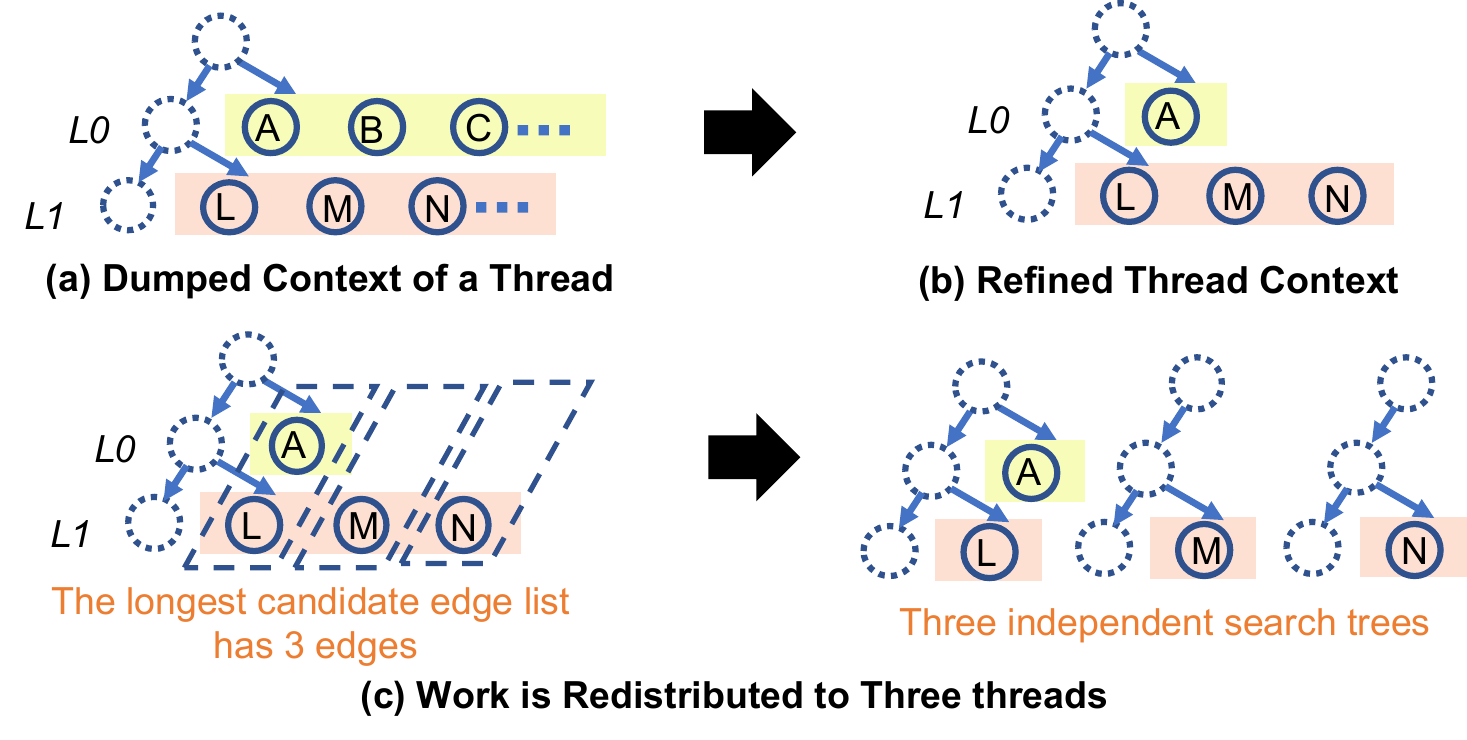}
  \vspace{-0.8cm}
  \caption{Illustration of tail warp work redistribution for a single thread.}
  \label{fig:tail-expand}
\end{figure}

\textit{Tail warps} are defined as a few warps with outstanding work when most other warps finish their execution and the task queues are empty.
This phenomenon exists because of the power-law nature of graph connectivity, where processing a small subset of search trees takes significantly longer time than most other trees.
Tail warps lead to load imbalance within a GPU kernel.
Consequently, the GPU baseline only achieves 22.8\% of occupancy and keeps 10 out of 48 warps active per SM (\S\ref{subsec:GPU_baseline_bottlenecks}).
The goal of tail warp work redistribution is to redistribute the work of the tail warps by exploiting sub-tree-level parallelism.

\THISWORK\ enables this optimization using a software work assignment mechanism.
Traditionally, GPU programs group parallel tasks using thread blocks and rely on hardware schedulers to assign work to SMs.
However, threads cannot inspect the execution status of the kernel in this model, because it is managed by the hardware.
This prevents warps from realizing whether they are tail warps or not.
In contrast, \THISWORK\ maintains a global work queue that dispatches work at warp granularity, which can be inspected to find the amount of outstanding work.
As the number of concurrently executing warps is limited, the global queue experiences minimal contention.
Using this mechanism, \THISWORK\ enables tail warp work redistribution in three subsequent steps as detailed below.





\textbf{Step 1. Abort tail warp execution.}
A warp aborts execution when the task queue is empty.
The first warp that exits signals all other warps to stop their current work via a variable.
Each warp checks the signaling variable periodically.
After receiving the signal, the threads in a warp set a timeout for its current execution.
If its work is not finished by the timeout, the thread dumps its context into a buffer, which includes candidate edge lists at all levels.
Empirically, \THISWORK\ reads the signaling variable every 1024 iterations and sets the timeout to 100 milliseconds.

\textbf{Step 2. Refine thread context.}
\THISWORK\ then post-processes the contexts of tail warps such that all edges in the candidate edge list are within the $\delta$-time window.
The invalid edges are removed by moving the end pointer of each candidate edge list forward.
The resulting edges satisfy temporal constraints and are most likely to spawn a sub-tree.
Similar to \S\ref{subsubsec:intra_warp_sharing}, the refined thread contexts can be shared by multiple threads to exploit STLP.
Fig.~\ref{fig:tail-expand}(a,b) illustrate this process with an example.
Before refinement, the thread context has two levels, each with a long candidate edge list.
After refinement, level 0 only has edge \texttt{A} and level 1 only contains edge \texttt{L}, \texttt{M}, and \texttt{N}.

\textbf{Step 3. Work redistribution.}
To complete the execution, \THISWORK\ launches a new kernel whose threads resume execution from the refined thread contexts.
This new kernel is treated as a regular kernel exhibiting all mechanisms including tail warp work redistribution.
If a refined context's longest candidate edge list has \texttt{N} edges, its candidate edges will be shared by \texttt{N} threads.
Fig.~\ref{fig:tail-expand}(c) illustrates how work is redistributed.
The refined thread context's longest candidate edge is at level 1, containing 3 edges.
Its candidate edges are redistributed and form three search trees, which are expanded independently.
The first tree contains edges \texttt{A} and \texttt{L}, and edge \texttt{M} and edge \texttt{N} are expanded by the other two trees.

\textbf{Theoretical Performance Analysis.}
The speedup introduced by this optimization can be approximated as follows:
$$
\frac{o}{1 - \Phi \mathcal{L}_\text{imb} } \leq o\phi \text{, where } \Phi = \frac{\phi - 1}{\phi} 
$$
We use $\phi$ to denote the total number of CUDA cores on a GPU divided by its warp width, and $\phi = 336$ on NVIDIA A40.
$o \lessapprox 1$ represents the small aggregated overhead of monitoring the signaling variable.
Let $0 < \mathcal{L}_\text{imb} < 1$ represent the fraction of total execution time consumed by a tail warp.
Note that $\mathcal{L}_\text{imb}$ can be significant even if the tail warp processes only a tiny fraction of the total work. 
When the tail warp is responsible for 1\% of the total work, $\mathcal{L}_\text{imb} \approx 0.77$.
When the work distribution is highly skewed, $\mathcal{L}_\text{imb}$ is large and this optimization can bring significant speedup (up to $o\phi$).
On the other hand, when $\mathcal{L}_\text{imb}$ is small, the overhead becomes noticeable.
When the tail warp's execution time is less than 100ms, this optimization does not take effect, and workload slows down by a factor of $o$.
The overhead is designed to be small, as the signaling mechanism only involves a few branches per iteration and one global memory access every 1024 iterations.

\subsection{Multi-GPU Support}
\subsubsection{Edge List Partitioning.} \label{subsubsec:edge_list_partitioning}
The goal of \THISWORK\ edge partitioning is to maximize efficiency by preventing inter-GPU communication.
Unlike prior graph partitioning techniques~\cite{metis}, \THISWORK\ does not use a convoluted partitioning scheme, because \THISWORK\ needs to partition graph edges in temporal order to ensure GPU-local accesses.
To this end, we define \textit{major and minor edge list partitions}.

If a search tree is expanded using a root node with an edge timestamp of $t_0$, and the coarse-grained time window is $\delta_0$, then this tree will only access edges within the time interval of [$t_0$, $t_0 + \delta_0$].
Therefore, only a small subset of edges need to be replicated on multiple GPUs at partition boundaries for all accesses to be GPU-local.
For an $N$-GPU system, the edge list is chronologically partitioned into $N$ contiguous sub-lists, called \textit{major partitions}.
A major partition is assigned to each GPU for processing.


Major partitions do not ensure all GPU-local accesses as trees expanded at partition boundaries span multiple major partitions.
Therefore, \THISWORK\ also generates \textit{minor partitions} for these edges on-the-fly after a query is provided to the system.
Assume a query with $\delta=\delta_0$ and a major partition contains all edges between the edges $e_i$ and $e_j$, where $i < j$.
Let $e_k$ be the last edge in this major partition such that $t_j - t_k > \delta_0$, where $t_k$ is the timestamp of $e_k$.
Edges in the range [$k+1$, $j$] are not mined in a major partition, as they cannot ensure GPU-local accesses.
Let $e_l$ be the first edge that satisfies $t_l - t_j > \delta_0$.
The corresponding minor partition includes edges within the range [$k + 1$, $l$], and can be used to mine patterns starting with edges in [$k+1$, $j$].
Although the construction of minor partitions is on the critical path of resolving a query, it does not bottleneck performance for two reasons.
First, queries often use small time windows, resulting in small minor partitions.
Second, we use GPUs to construct minor partitions, which takes negligible time compared to the mining itself.

\subsubsection{Multi-GPU Scheduling.} \label{subsubsec:mgpu-sched}
To reduce the work imbalance among multiple GPUs, \THISWORK\ runtime dynamically schedules work on multiple GPUs.
This mechanism is only enabled for major partitions as minor partitions only use a small amount of time.
Each major partition is divided into 16 sub-partitions and placed into the task queue.
Each GPU gets a range of edges to mine from its task queue.
In this scheme, an idle GPU inspects the work queues of other busy GPUs, and identifies the queue with the highest amount of outstanding work.
Once one is identified, the idle GPU steals work from the busy GPU at a sub-partition granularity.
This mechanism continues until all GPU work queues are drained.

\subsection{Support for Additional Constraints}
\subsubsection{Vertex/Edge Labels.}
\THISWORK\ code generator generates functions that examine vertex and edge labels based on the user-defined query.
Whenever a new vertex/edge is matched, the generated functions are used to check their validity.

\subsubsection{Fine-Grained Temporal Constraints.}
The fine-grained temporal constraints impose unique conditions over each level of the search tree.
The time constraint is computed after an edge is matched and recovered after a backtrack through re-computation.

\subsubsection{Temporal Anti Edges.}
During search tree expansion, temporal anti-edges are expanded in a similar fashion as a normal edge.
In contrast to a motif edge, a presence/match of an anti-edge results in the termination of a tree branch.
When an anti-edge is specified with fine-grained temporal constraints, the code generator generates an array whose $i^{th}$ entry contains the last real edge before a valid edge $i$.
This array will be referred in order to compute a correct time limit.
The work in the levels corresponding to anti-edges cannot be shared through intra-warp work stealing or tail warp work redistribution.
These levels do not produce candidate edges that spawn sub-trees, which makes the two optimizations based on STLP inapplicable.



\section{Evaluation methodology} \label{sec:methodology}
\textbf{Input Temporal Graphs.}
\begin{table}[t]
  \centering
  \scriptsize 
  \begin{tabular}{ c | r | r | r | c}
  \hline
  \rule{0pt}{10pt} 
  \textbf{Graph} & \textbf{\#Vertices} & \textbf{\#Temporal}  &   \textbf{\# Static}      & \textbf{Time span} \\ & & \textbf{Edges} & \textbf{Undirected Edges} &  \textbf{(years)} 
  \tabularnewline
  \hline
  \hline
  wikitalk (wi)~\cite{snap} & 1,140,149 & 7,833,140 & 2,787,968 & 6.24 \\
  stackoverflow (so)~\cite{snap} & 2,601,977 & 63,497,050 & 34,875,685 & 7.6 \\
  reddit-reply (re)~\cite{liu2018sampling} & 8,901,033 & 646,044,687 & 435,290,421 & 10.1 \\
  ethereum (eth)~\cite{ethdatasets} & 66,323,478 & 628,810,973 & 186,064,655 & 3.58 \\
  \hline
  \end{tabular}
  \caption{Temporal graph datasets used for evaluation.}
   \vspace{-12pt}
  \label{table:input_dataset}
\end{table}
To evaluate the performance of \THISWORK, we use four real-world temporal graphs as shown in Table~\ref{table:input_dataset}.
The chosen datasets are among the largest found in the public domain, exhibiting diversity in terms of their sizes, connectivity, and granularity of temporal interactions.
\textllsm{wiki-talk} represents Talk page editing activities among Wikipedia users.
\textllsm{stackoverflow} includes interactions between users on Stack Overflow.
\textllsm{temporal-reddit-reply} encompasses reddit replies.
\textllsm{ethereum} is a transaction network extracted from Ethereum. 
Given that graphs are non-attributed, we attach random node and edge features.

\noindent
\textbf{User-Defined Queries.}
\begin{figure}[t]
    \centering
    \includegraphics[width=\linewidth]{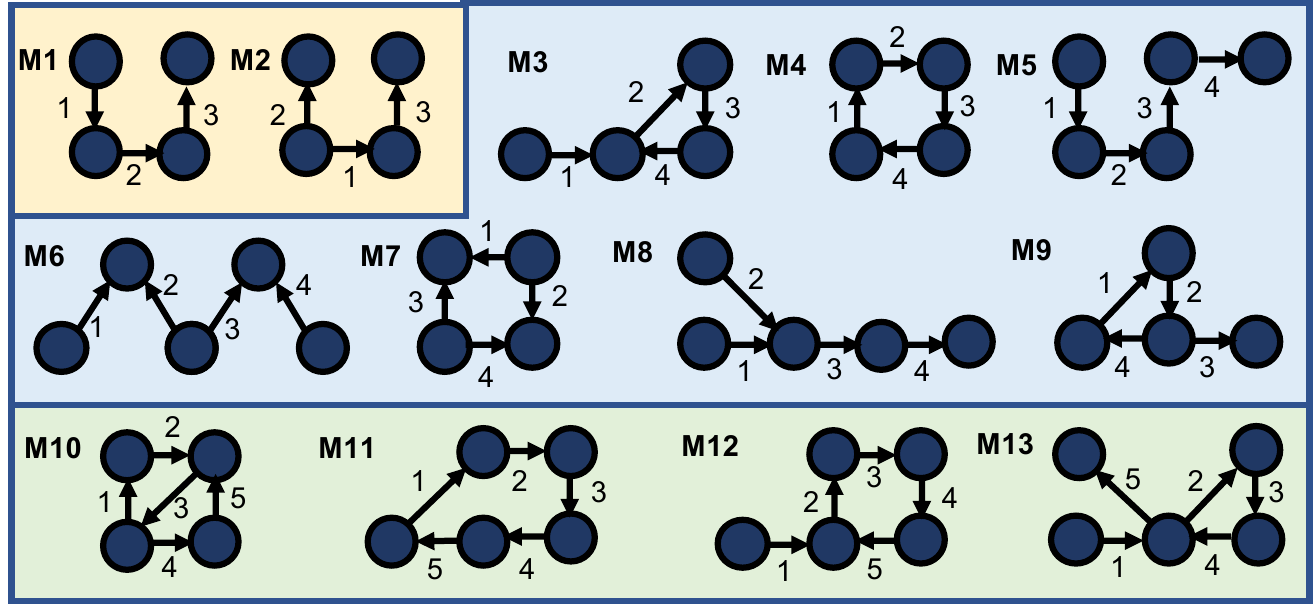}
    \vspace{-0.8cm}
    \caption{Temporal motifs used for evaluation.}
    \vspace{-0.35cm}
    \label{fig:motifs}
\end{figure}
We select 13 distinct motifs from prior works~\cite{mackey2018chronological,sarpe2021presto,Kosyfaki2018FlowMI} with three to five edges in size, as depicted in Fig.~\ref{fig:motifs}. 
We only use the pattern and edge order of motifs from~\cite{Kosyfaki2018FlowMI}, ignoring flow-based matching rules.
This set of motifs enables a comprehensive evaluation on our system's performance. 
Since the difficulty of mining specific motif patterns varies across different target graphs, we employ distinct time windows for each graph due to limited time for experiments.
Regardless of additional constraints, we set use $\delta=1$ day for \textllsm{wiki-talk} and \textllsm{stackoverflow}, $\delta=10$ hours for \textllsm{temporal-reddit-reply} and $\delta=1$ hour for \textllsm{ethereum}.
We use \textllsm{wiki-talk} and motif \texttt{M6} for all profiling activities. 

\noindent
\textbf{Hardware Platform Configuration.}
We run CPU baselines on a dual-socket server with two Intel Xeon Platinum 8380 processors, each with 40 physical cores (80 SMT threads) and 8 memory channels.
The main memory capacity is 1TB.
We use up to four NVIDIA A40 GPUs to evaluate our design, each with 48GB GDDR6 memory.
We use Nsight Compute to collect profiling results.






\section{Evaluation Results} \label{sec:results}



\textbf{Effectiveness of \THISWORK\ Optimizations.}
\begin{figure*}[t]
\centerline{\includegraphics[width=\linewidth]{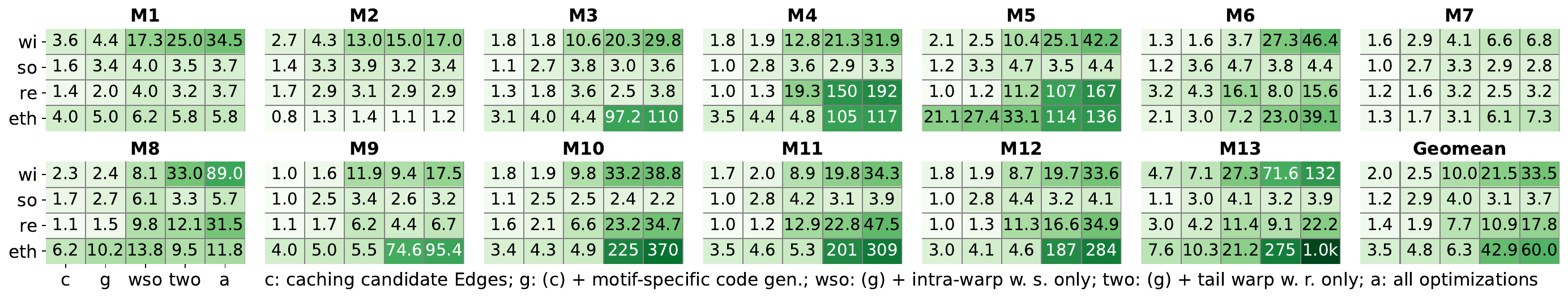}}
\vspace{-0.4cm}
\caption{Performance improvements ($\times$) of \THISWORK\ with different optimizations compared to the GPU baseline (\S\ref{subsec:GPU_baseline_bottlenecks}).}
\vspace{-0.35cm}
\label{fig:heatmap}
\end{figure*}
Fig.~\ref{fig:heatmap} shows the improvements of \THISWORK\ performance with various combinations of optimizations enabled, normalized to the GPU baseline discussed in \S\ref{subsec:GPU_baseline_bottlenecks}.
The figure shows that all proposed optimizations (last column \texttt{a} of the heatmap) improve the performance by 3.7$\times$--60$\times$ (19$\times$), on average for different input graphs.
These results are based on single-GPU experiments.
These significant performance improvements are attributed to improved per-thread algorithmic efficiency, reduced memory footprint, and intelligent load balancing techniques employed in \THISWORK.
\THISWORK\ system design hides all the implementation details of sophisticated optimizations from the programmer; the user is only expected to set up a query of interest and simple runtime parameters (Fig.~\ref{fig:query_example}).

The figure further shows that caching candidate edges (column \texttt{c}) and generating query-specific mining plan (column \texttt{g}) improves the performance of the baseline GPU by 1.2$\times$--4.8$\times$, on average.
These optimizations improve the per-thread algorithmic work performed for mining each motif, \textit{i.e.,} by reducing the number of costly binary search and control flow operations that depend on dynamic runtime information.
Adding intra-warp work stealing (column \texttt{wso}) further improves performance by 4$\times$--10$\times$.
This mechanism balances work distribution among multiple threads in the same warp by the proposed employing sub-tree-level parallelism.
Another work balancing mechanism, \textit{i.e.,} tail warp work redistribution (column \texttt{two}) shows a more pronounced effect on performance with an uplift of 3.1$\times$--42.9$\times$.
This optimization redistributes work from large search trees that run for the highest amount of time.
A few long-running warps under-utilize GPU hardware resources as most other warps remain idle.
Employing tail warp work redistribution enables \THISWORK\ to exploit full potential of the GPU hardware resources, resulting in a significant performance uplift.
These results also emphasize a heavy work imbalance while running this workload, underscoring the significance of our optimizations.

By comparing column \texttt{g} and \texttt{two} in \texttt{M2}, \texttt{M8}, and \texttt{M10}, note that applying tail warp work redistribution occasionally slows the execution down marginally.
This is because they do not trigger this optimization due to short tail warps, and only pay for the overhead of signal monitoring as a result.
More frequent slowdowns are observed when this optimization is applied after the application of intra-warp work stealing (columns \texttt{wso} and \texttt{a}).
This is because intra-warp work stealing can shorten long tail warps, making tail warp work redistribution less likely to happen.
This small overhead varies for different inputs due to the different micro-architecture resource requirements.
Such minor slowdowns do not dim the overall effectiveness of tail warp work redistribution, as the speedups in the most-common highly-skewed cases necessitate its existence.

Interestingly, the effectiveness of \THISWORK\ optimizations can be vastly different for mining different motifs from graphs.
Take an example of the largest \textllsm{ethereum} input graph for example (row \texttt{eth} in different heatmaps).
The performance improvement brought by \THISWORK\ ranges from 1.2$\times$ (\texttt{M9}) to 1022.1$\times$ (\texttt{M13}).
Mining \texttt{M4} achieves high efficiency due to load balancing mechanisms, while \texttt{M5} takes more advantage over the GPU-friendly implementation.
Overall, the motifs with more edges are more likely to induce load imbalance and emphasize the importance of load balancing mechanisms, and \THISWORK\ consistently delivers performance improvements.

\textbf{Detailed Performance Analysis.}
\begin{figure}[t]
\centerline{\includegraphics[width=\linewidth]{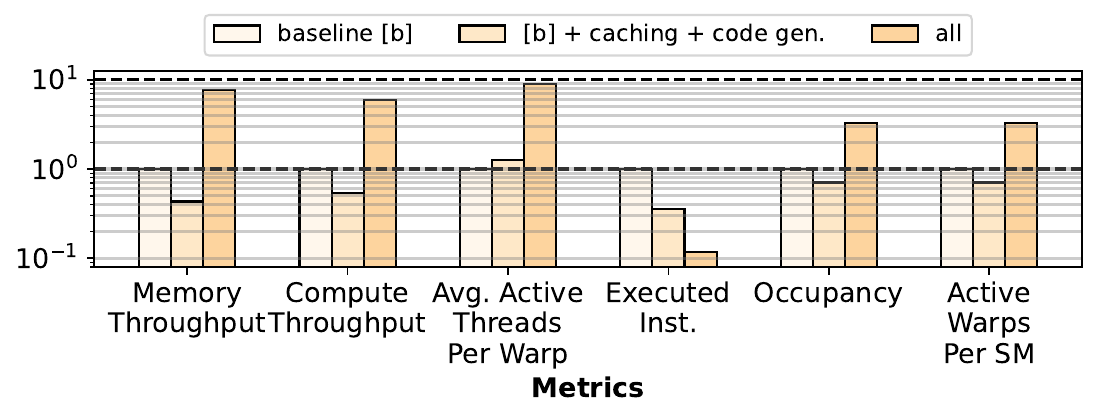}}
\vspace{-0.5cm}
\caption{Detailed performance analysis for a representative mining of \texttt{M6} on \textllsm{wiki-talk}, relative to the GPU baseline.}
\vspace{-0.5cm}
\label{fig:cmp_prof}
\end{figure}
Fig.~\ref{fig:cmp_prof} compares detailed performance metrics outputted by the GPU profiler for 1) GPU baseline, 2) thread-local optimizations, and 3) all optimizations that also include load balancing.
The thread-local optimizations reduce the total number of executed instructions by 64\%.
This further verifies the reduction in costly binary search operations.
On the other hand, load-balancing mechanisms improve resource utilization on GPUs at all levels.
These increase the number of active threads per warp by 8.9$\times$ and improve occupancy and warps per SM by 3.3$\times$.
As a result, the compute throughput and memory throughput are improved by 6.0$\times$ and 7.8$\times$.
Although the load balancing at runtime uses GPU instructions, the final effect of this optimization reduces the number of execution instructions overall.
In particular, \THISWORK\ only executes 11.7\% of the instructions compared to the baseline.
This is because GPU follows Single Instructions Multiple Thread (SIMT) execution model, where one instruction is issued for multiple threads in a warp.
As the number of active threads in a warp increases as a result of load balancing, more threads share the same instruction, reducing the total instruction count.

\textbf{Comparison with a CPU Baseline.}
\begin{table}[t]
  \centering
  \scriptsize 
  \begin{tabular}{ c | r | r | r | r | r | r | r | r | r }
  \cline{2-10}
  \multicolumn{1}{c}{} & \multicolumn{3}{c|} {\textbf{stackoverflow}}& \multicolumn{3}{c|}{\textbf{reddit-reply}}& \multicolumn{3}{c}{\textbf{ethereum}} \\
  \hline
  \textbf{Impl.} & \textbf{\texttt{M5}} & \textbf{\texttt{M6}} & \textbf{\texttt{M7}} & \textbf{\texttt{M5}} & \textbf{\texttt{M6}} & \textbf{\texttt{M7}}  & \textbf{\texttt{M5}} & \textbf{\texttt{M6}} & \textbf{\texttt{M7}}
  \tabularnewline
  \hline
  \hline
  CPU BL & 7.4 & 12.6 & 6.6  & 1247.9 & >10 hr & 2895.5 & 929.6 & >1 day  & 16802.3 \\
  GPU BL & 1.0 & 2.1 & 0.9  & 2920.0 & 7952.1 & 474.6 & 300.6 & 67099.1 & 5273.1  \\
  \THISWORK\ & 0.2 & 0.5 & 0.3 & 17.2 & 510.2 & 148.8 & 2.2 & 1717.1 & 726.7  \\
  \hline
  \textbf{Impl.} & \textbf{\texttt{M8}} & \textbf{\texttt{M11}} & \textbf{\texttt{M12}} & \textbf{\texttt{M8}} & \textbf{\texttt{M11}} & \textbf{\texttt{M12}}  & \textbf{\texttt{M8}} & \textbf{\texttt{M11}} & \textbf{\texttt{M12}} 
  \tabularnewline
  \hline
  \hline
  CPU BL & 9.3 & 8.6 & 8.7  & 6031.1 & 3881.8 & 13493.0  & >10 hr & 33024.6 & 32812.6 \\
  GPU BL & 2.2 & 1.7 & 1.7  & 3704.8 & 3105.2 & 3118.6    & 7856.5 & 11921.2 & 12843.8   \\
  \THISWORK\ & 0.4 & 0.4 & 0.4   &  117.6 & 64.0 & 90.0  & 665.6 & 38.6 &  45.2  \\
  \hline
  \end{tabular}
  \caption{Execution time comparison (sec). GPU BL: GPU Baseline; CPU BL: CPU Baseline based on Mackey \textit{et al.}~\cite{mackey2018chronological}. } 
   \vspace{-0.75cm}
  \label{table:cmp-impl}
\end{table}
Next, we compare the runtime of long-running motifs on the three large input graphs (due to space limitation) for a baseline CPU and GPU baselines (based on Mackey \textit{et al.}~\cite{mackey2018chronological}), and \THISWORK.
The CPU baseline uses an in-house parallel version of the open-source implementation of Mackey \textit{et al.}'s algorithm with 64 threads.
Table~\ref{table:cmp-impl} compares the absolute mining time in seconds for three baselines.
\THISWORK\ outperforms the state-of-the-art CPU baseline by 62.1$\times$, on average.
This table shows that merely transforming CPU code into GPU code is not enough as the baseline CPU can outperform GPU in some cases (\textit{e.g.,} the CPU baseline outperforms the GPU baseline by 2.3$\times$ when mining \texttt{M5} on \textllsm{reddit-reply}).
This is especially true for irregular workloads like graph mining that suffer from long memory latency, frequent thread divergence, and load imbalance.
Table~\ref{table:cmp-impl} emphasizes the \THISWORK\ optimization effort that significantly improves workload performance.
For example, \THISWORK\ can mine \texttt{M6} in less than 30 minutes, versus >18 hours for a GPU baseline.


\textbf{Scaling with Multiple GPUs.}
\begin{figure}[t]
\centerline{\includegraphics[width=\linewidth]{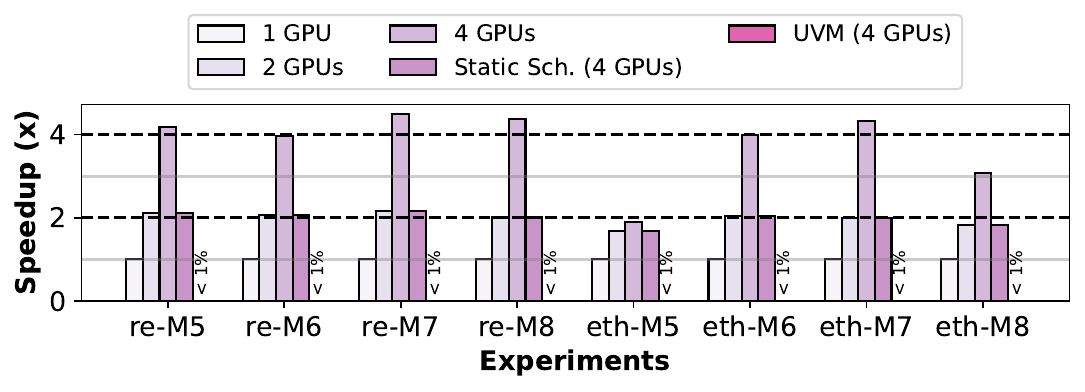}}
\vspace{-0.5cm}
\caption{Effectiveness of multi-GPU support. \textllsm{re-M6} means mining \texttt{M6} on \textllsm{reddit-reply}. Other experiments follow the same naming convention.}
\vspace{-0.5cm}
\label{fig:multi_gpu_perf}
\end{figure}
To study the performance scaling with multiple GPUs, we compare the performance of \THISWORK\ for different number of GPUs that mine \texttt{M5}-\texttt{M8} in \textllsm{reddit} and \textllsm{ethereum}.
Fig.~\ref{fig:multi_gpu_perf} shows that the performance of \THISWORK\ scales well with multiple GPUs for most workloads.
The speedup numbers can sometimes surpass the number of GPUs due to the side effect of partitioning the edge list.
This is because it is cheaper to generate candidate edge lists using a partition than using the whole graph as there are fewer visible neighbors to search for a given vertex.
Neglecting this effect, \THISWORK\ scales linearly with the number of GPUs for motifs \texttt{M6} - \texttt{M8} on both graphs.
One exception to this trend is mining \texttt{M5} on the \textllsm{ethereum} graph, where the proposed work partitioning scheme results in a disproportionately high amount of work assigned to one of the GPUs due to the large population of motif matches in one of the partitions.
Furthermore, this workload takes only 2.2s (Table~\ref{table:cmp-impl}) to run on a single-GPU, leaving very little work for a multi-GPU system to scale.

We compare \THISWORK\ with two additional implementations to showcase the benefits of multi-GPU optimization strategies.
If we statically assign each major partition to a GPU without using our dynamic scheduling mechanism, the system's performance stops scaling beyond 2$\times$, showcasing the effectiveness of the proposed dynamic scheduling.
Without partitioning graphs, graphs have to be replicated on multiple GPUs, preventing them from processing large graphs.
GPUs can oversubscribe GPU memory via Unified Virtual Memory (UVM), however, it only delivers $< 1$\% performance compared to a single GPU due to the overhead of demand paging~\cite{NVBlog}.
This shows the effectiveness of proposed major/minor partitions.

\textbf{Incorporating Additional Constraints.}
\begin{table}[t]
\centering
\scriptsize
\begin{tabular}{ c | r | r | r | r | r | r | r | r }
\cline{2-9}
\multicolumn{1}{c}{} & \multicolumn{2}{c|} {\textbf{wiki-talk}} & \multicolumn{2}{c|}{\textbf{stackoverflow}} & \multicolumn{2}{c|}{\textbf{reddit-reply}} & \multicolumn{2}{c}{\textbf{ethereum}} \\
\hline
\textbf{Cons.} & \textbf{matches} & \textbf{time} & \textbf{matches} & \textbf{time} & \textbf{matches} & \textbf{time} & \textbf{matches} & \textbf{time}
\tabularnewline
\hline
N & 1.5e5 & 20 & 5.0e5 & 215 & 7.4e8 & 13868 & 6.4e9 & 2348 \\
V & 8.8e3 & 8 & 3.1e4 & 50 & 3.2e7 & 1620 & 3.4e5 & 367 \\
V+T & 2.0e3 & 6 & 1.4e4 & 36 & 1.7e7 & 916 & 9.2e4 & 198 \\
V+T+A & 1.5e3 & 7 & 1.2e4 & 44 & 6.3e6 & 1076 & 8.7e4 & 254 \\

\hline
\end{tabular}
\vspace{-0.05cm}
\caption{Effect of incorporating additional mining constraints on the number of matches and execution time (in milliseconds) for 4-cycle. N: No constraints; V: vertex labels; T: fine-grained temporal constraints; A: anti-edge, additive.}
\vspace{-1cm}
\label{table:additional-features}
\end{table}
Table~\ref{table:additional-features} demonstrates the effects of applying vertex labels, fine-grained temporal constraints, and anti-edges to mine temporal 4-cycles.
After applied all the constraints, the query is similar to the one shown in Fig.~\ref{fig:query_example}.
As the labels are randomly generated, we only conduct experiments for vertex labeling without loss of generality.
Both vertex labels and additional temporal constraints reduce the number of valid matches and execution time by orders of magnitude.
Anti-edge constraints, on the other hand, results in a reduced matches but increased execution time due to overhead in verifying this constraint.

\textbf{Comparison with Static Graph Mining Algorithms.}
One valid way to mine temporal motifs is to first mine static motifs, and then resolve temporal constraints~\cite{paranjape2017motifs}.
This however, often leads to significantly more amount of algorithmic work as many valid static motifs do not follow temporal constraints.
Table~\ref{table:vs-static} presents the ratio of the number of static to temporal motifs matched, which clearly shows that static mining algorithms perform large amounts of unnecessary work.
When multiple temporal motifs map to the same static motif, we present the ratio with the lowest motif ID.
This is the reason to base \THISWORK\ on Mackey \textit{et al.}'s algorithm that resolves both structural and temporal constraints together.

\section{Related Works}

\textbf{Software Frameworks for Temporal Motif Mining.}
Several software frameworks are proposed for temporal motif mining, as discussed in \S\ref{subsec:prior-work}.
Most of them aim at finding $\delta$-window temporal motifs. 
Algorithms proposed in~\cite{paranjape2017motifs,mackey2018chronological,kumar20182scent} target exactly matching motifs from graphs, while the ones from~\cite{wang2020efficient,liu2018sampling,sarpe2021presto} approximately mining motifs by sampling a subset of edges to achieve better scalability.
A few other works~\cite{kovanen2011temporal, pasha21Kdd} adopt a more general problem definition that considers the timestamp difference between consecutive edges.
All of them are CPU-based solutions that focus on a specific temporal motif setting.
In contrast, \THISWORK\ is a high-performance GPU-based system that can process a wide range of temporal motifs via queries.

\begin{table}[t]
  \centering
  \scriptsize 
  \begin{tabular}{ c | r | r | r | r }
  \hline
  \textbf{Static (Temporal) Motif} & \textbf{wiki-talk} & \textbf{stackoverflow} & \textbf{reddit-reply} & \textbf{ethereum}
  \tabularnewline
  \hline
    3-path (\texttt{M1}) & 9.6e6 & 5.7e8 & 9.8e5 & 2.7e3 \\
    tailed-tri (\texttt{M3}) & 1.1e5 & 7.1e5 & 5.6e4 & 1.5e4 \\
    fourcycle (\texttt{M4}) & 1.6e4 & 9.2e4 & 1.3e4 & 1.7e2 \\
    static M8 (\texttt{M8}) & 1.4e8 & 1.5e8 & 9.5e6 & 2.2e6 \\
    diamonds (\texttt{M10}) & 1.9e4 & 9.6e5 &  1.3e3  & 1.8 \\
    static M13 (\texttt{M13}) & 4.8e10 & 1.3e7 & >9.5e11 & 3.7e6 \\
  \hline
  \end{tabular}
  \caption{Ratio of the number of matches generated by static graph mining algorithms versus temporal motif mining.
  }
   \vspace{-0.9cm}
  \label{table:vs-static}
\end{table}

\noindent
\textbf{Software Frameworks for Static Graph Mining.}
Multiple systems are proposed for static graph pattern mining~\cite{arabesque,peregrine,peregrine_followup,graphzero,automine,graphpi,scalable22icde,oden21, alex22analytical,stmatch22, G2Miner2022OSDI}.
The problem of temporal motif mining adds temporal constraints to the static mining problem, which none of these prior systems can support.
CPU execution is more tolerant to load imbalance and complex control flow due to fewer cores and sophisticated branch prediction hardware.
While~\cite{G2Miner2022OSDI} performs mining code generation on GPUs, temporal motif mining imposes additional challenges.
Utilizing temporal motifs increases programs' control flow complexity and per-thread resource usage are unfavorable for GPU execution and require a carefully designed load-balancing mechanism.
\THISWORK\ is the first system that generates high-performance GPU code for temporal motif mining.

\section{Conclusion}
This paper presented \THISWORK---an optimized system for mapping the workload of temporal motif mining onto the GPU architecture.
Our study showed that existing software for executing this workload yields poor performance on a GPU due to long-latency memory instructions, frequent thread divergence operations, and heavy load imbalance.
\THISWORK\ presented domain-specific optimizations to address these inefficiencies.
In particular, we showed how \THISWORK\ generates per-thread code using motif structure and temporal constraints to cache key metadata information to reduce costly memory and thread divergent operations.
Furthermore, we presented \THISWORK\ runtime primitives that enable load balancing to improve the GPU hardware utilization.
For large input graphs that do not fit in the GPU memory, \THISWORK\ employs low-cost edge list partitioning that prevents inter-GPU communication.
\THISWORK\ is easy to use, where a targeted user only needs to write an input query, and the heavy lifting of code generation is automatic by the system.
We showed that compared to a baseline GPU implementation, \THISWORK\ uplifts the performance of 19$\times$, on average.
While \THISWORK\ is designed for a specific workload, its design philosophy will inspire future research on efficiently mapping data-intensive irregular workloads onto GPU hardware.



\begin{acks}
This research is based upon work supported by the Office of the Director of National Intelligence (ODNI), Intelligence Advanced Research Projects Activity (IARPA), through the Advanced Graphical Intelligence Logical Computing Environment (AGILE) research program, under Army Research Office (ARO) contract number W911NF22C0085. The views and conclusions contained herein are those of the authors and should not be interpreted as necessarily representing the official policies or endorsements, either expressed or implied, of the ODNI, IARPA, ARO, or the U.S.  Government. This work was also supported by the United States-Israel BSF grant number 2020135.
\end{acks}

\newpage

\appendix
\section{Theoretical Performance Analysis for Load Balancing Mechanisms} \label{app:model}
\textbf{Intra-warp Work Stealing:}
Each trigger of this optimization uses around 25 fast warp-level primitives~\cite{NVBlogb} for work stealing.
Because these warp-level primitives use GPU registers and do not introduce complicated control flow, their aggregated cost is far less than a normal search iteration's cost. 
The warp-level speedup of this optimization can be roughly modeled as follows:
$$
\frac{\mathcal{I}_\text{imb}}{\mathcal{I}_\text{opt} + k \epsilon} = \frac{\mathcal{I}_\text{imb}}{\mathcal{I}_\text{opt}} \frac{\mathcal{I}_\text{opt}}{\mathcal{I}_\text{opt} + k \epsilon} = \frac{32}{t_\text{imb}} \frac{\mathcal{I}_\text{opt}}{\mathcal{I}_\text{opt} + k \epsilon} = \frac{32}{ t_\text{imb} \left( 1 + \frac{k\epsilon}{\mathcal{I}_\text{opt}} \right) }
$$
$\mathcal{I}_\text{imb}$ and $\mathcal{I}_\text{opt}$ are the number of search iterations used by the imbalanced baseline and optimized code.
$k < \mathcal{I}_\text{opt}$ is the times this mechanism is triggered and $\epsilon < 1$ is the work-stealing overhead relative to a search iteration. 
$t_\text{imb}$ is the number of active threads within a warp in the imbalanced baseline.  

We observe that $t_{\text{imb}}$ is usually much smaller than warp width 32 due to intra-warp work imbalance.
Each trigger of this optimization steals multiple sub-trees, resulting in a smaller $k$.
Our warp-primitive-based efficient implementation results in a smaller $\epsilon$.
In practice, we observe $t_\text{imb} \left( 1 + \frac{k\epsilon}{\mathcal{I}_\text{opt}} \right) < 32$, and a meaningful speedup on average across warps.
In theory, when the work within a warp is naturally balanced, this optimization can slightly slow down the execution by a factor of $\frac{k \epsilon}{\mathcal{I}_\text{opt}}$.
However, as most of the large graphs are power-law graphs, it is unlikely to happen in practice. 
A trivial exception is all threads in a warp do not have much work, so we enable this optimization after 20 iterations to filter out these warps. 
Another interesting fact is that the speedup is  correlated to $\mathcal{I}_\text{opt}$.
In other words, the more work is assigned to a warp, the more speedup it can enjoy from this optimization.
This observation helps explain the interaction between this optimization and tail warp work redistribution mechanism.

For simplicity, we ignore the negligible overhead of detecting idle threads within warps, which additionally uses a warp vote and a branch in each search iteration.
Also, we assume that all threads can be busy after this optimization is enabled.
This is the usual case except near the end of a warp's execution, where there is not enough work for 32 threads to do. 

The above model explains the benefit of this optimization at the warp level, while it is overly complicated to infer the benefit for the entire kernel from this model.
Factors like hardware schedulers' algorithms and work distribution are either not released to the public or are dependent on the user's input.

\noindent \textbf{Tail Warp Work Redistribution:}
For simplicity, we assume that the imbalanced work distribution causes one tail warp keeps running while all others finish. 
Our analysis can be generalized to cases with multiple tail warps.
We first model the negative impact of tail warps.
Let $0 < \mathcal{L}_\text{imb} < 1$ represent the fraction of total execution time consumed by a tail warp.
Note that $\mathcal{L}_\text{imb}$ can be significant even if the tail warp only processes a tiny fraction of all work. 
We use $\phi$ to represent the total number of CUDA  cores on a GPU divided by its warp width, and $\phi = 336$ on NVIDIA A40.
Assume that one warp near the end of the execution handles 1\% of all work, which is reasonable based on our profiling result as shown in Figure 3(a).
In such case, $\mathcal{L}_\text{imb} = \frac{1}{(99/336) + 1} \approx 77\%$, and GPU is severely underutilized in most of its execution time.
Such, or even worse, scenario frequently happens in practice and is hard to foresee as it depends on graphs, motifs, and constraints.

Then, we model the benefit of our tail warp work redistribution optimization.
Let $c$ represent the cost to abort execution, refine thread context, and redistribute the work via kernel launch, which is around several microseconds.
We use $k$ to represent the times this optimization is triggered, and $T_\text{imb}$ to model the total execution time.
We use $o \lessapprox 1$ to represent the small aggregated overhead of monitoring the signaling variable.
The overhead is designed to be small, as the signaling mechanism only involves a few branches per iteration and one global memory access every 64 iterations.
Note that the overhead also weakly depends on mining queries.
Different queries stress micro-architecture resources differently, resulting in different degrees of resource competition between normal execution and signal monitoring.
The speedup introduced by this optimization can be roughly modeled as follows:
$$
\frac{T_\text{imb}}{(1 - \mathcal{L}_\text{imb} + \frac{\mathcal{L}_\text{imb}}{\phi}) T_\text{imb} + kc} \cdot o  = \frac{o}{(1 - \frac{\phi - 1}{\phi} \mathcal{L}_\text{imb} ) + \frac{kc}{T_\text{imb}}}
$$

Usually, a couple of times work redistribution is enough, so $\frac{kc}{T_\text{imb}}$ is small enough to ignore.
The speedup is further approximated by
$$
\frac{o}{1 - \frac{\phi - 1}{\phi} \mathcal{L}_\text{imb} } = \frac{o}{1 - \Phi \mathcal{L}_\text{imb} } \leq o\phi \text{, where } \Phi = \frac{\phi - 1}{\phi} 
$$

When the work distribution is highly skewed, $\mathcal{L}_\text{imb}$ is large and this optimization can bring significant speedup (up to $o\phi$).
On the other hand, when $\mathcal{L}_\text{imb}$ is small, the overhead becomes noticeable.
When the tail warp's execution time is less than 100ms, this optimization does not take effect, and workload slows down by a factor of $o$.

\noindent \textbf{Interaction between Optimizations:}
Applying intra-warp work stealing can decrease $\mathcal{L}_\text{imb}$.
Tail warps are the warps that are assigned with large amounts of work.
As demonstrated by the performance model, such kinds of warps are privileged with regard to intra-warp work stealing.
Plus, many warps with little work do not trigger intra-warp work stealing due to the threshold. 
There is a net difference between the speedup gained from intra-warp work stealing for tail warps and normal warps.
If we use $\theta > 1$ to represent the ratio between these two values, then intra-warp work stealing decreases $\mathcal{L}_\text{imb}$ to 
$$
\mathcal{L}_\text{imb}^\prime = \frac{\frac{\mathcal{L}_\text{imb}}{\theta}}{1 -  \mathcal{L}_\text{imb} + \frac{\mathcal{L}_\text{imb}}{\theta} }  = \frac{1}{\theta(1 - \mathcal{L}_\text{imb}) + \mathcal{L}_\text{imb}} \cdot \mathcal{L}_\text{imb} < \mathcal{L}_\text{imb}
$$
Effectively, intra-warp work stealing eases the tail warps problem.

As a result, applying tail warp work redistribution upon intra-warp work stealing results in a smaller speedup number compared to the case without intra-warp work stealing results.
When the work distribution is only slightly skewed, intra-warp work stealing is good enough to tame the tail warps problem.
In such cases, the tail warp's execution time is less than 100ms, and tail warp work redistribution exhibits a factor of $o$ slowdown.

\section{Qualitative Comparison with Closely Related Works} \label{app:comparison}
Next, we discuss the unique features of \THISWORK\ system design compared to closely related prior works~\cite{emptyhead17sigmod, graphzero, automine, graphpi,G2Miner2022OSDI,pangolin,stmatch22}.

\textbf{Code Generation Strategy.}
Many prior graph mining systems adopt the idea of code generation~\cite{emptyhead17sigmod, graphzero, automine, graphpi, G2Miner2022OSDI}, for CPUs or GPUs.
Nevertheless, none of them takes temporal constraints into consideration, and can only generate code to mine static patterns.
While CPU-based code generation has been done by~\cite{emptyhead17sigmod, graphzero, automine, graphpi}, CPU execution is more tolerant to load imbalance and complex control flow due to fewer cores and sophisticated branch prediction hardware.
Because of the complexity of GPU programming, only~\cite{G2Miner2022OSDI} has done GPU code generation for static pattern mining.
However, its strategy falls short in the temporal settings, because mining temporal motifs on GPU poses unique challenges compared to mining static motifs on GPU. 
Utilizing temporal constraints increases programs’ control flow complexity and per-thread resource usage, which are unfavorable for GPU execution and require a carefully designed load-balancing mechanism. 
The lack of parallelizable operations, like vertex set operations in~\cite{G2Miner2022OSDI}, makes it infeasible to utilize warps via SIMD-style programming.
This requires the code generator to generate highly efficient thread-granularity load balancing code that is precisely based on resource allocation status.
Therefore, unlike~\cite{G2Miner2022OSDI}, Everest’s code generator focuses on reducing control flow complexity, minimizing resource usage, and is coupled with load-balancing mechanisms.
These observations are shown to be the key to good performance for temporal motif mining, which are not addressed in~\cite{G2Miner2022OSDI}'s code generator for GPUs.

\textbf{Load Balancing Techniques.}
While GPU-based graph mining systems can achieve a balanced work distribution using BFS search order~\cite{pangolin}, it is at the expense of excessive memory usage, which is not acceptable for large graphs on GPUs. 
Many high-performance GPU-based graph mining systems thus switch to DFS search order~\cite{G2Miner2022OSDI, stmatch22}, and propose their load balancing optimizations for skewed work distribution.
These methods differ significantly from the ones in \THISWORK.
G2Miner~\cite{G2Miner2022OSDI} only has one load balancing mechanism for single GPU execution, which neither uses work stealing nor is applicable to all patterns.
STMatch~\cite{stmatch22} does not redistribute work for tail warps and its intra-block/grid work stealing mechanisms are at the expense of maintaining complicated data structures in memory. 
Both~\cite{G2Miner2022OSDI, stmatch22} use SIMD vertex set operation primitives to fully utilize warps.
These methods are not applicable in temporal settings, due to the differences in workload characteristics between DFS-based temporal motif mining and static pattern mining. 
First, search trees in temporal settings are expanded by mapping new edges and do not require vertex set operations or anything similar, thus warps cannot be efficiently utilized via SIMD-style programming. 
Second, the time constraints on motifs shorten most of the search trees, make warp-granularity execution~\cite{stmatch22, G2Miner2022OSDI} inefficient, and call for thread-granularity load balancing mechanisms like the ones in \THISWORK. 
Third, the neighborhood information depends on search tree roots due to temporal constraints, which vagues the remaining work in search trees and forbids high-overhead work stealing mechanisms that rely on accurate work representations~\cite{stmatch22}. 
As a result, \THISWORK\ proposes two novel load balancing solutions, intra-warp load balancing and tail-warp work redistribution, without maintaining expensive data structures or defining new SIMD operation primitives.
Intra-warp load balancing can do work stealing at thread granularity only using registers, in contrast to the intra-block solution in~\cite{stmatch22} that only works at warp granularity and has to use shared memory.
Tail-warp work redistribution relies only on a signaling variable to redistribute work among the whole GPU, while the intra-grid solution in~\cite{stmatch22} maintains a complicated data structure in global memory for remaining work and requires busy waiting on locks for work stealing.

\textbf{Memoization Technique.} 
While, on the surface, the caching mechanisms of \THISWORK\ and STMatch~\cite{stmatch22} sound similar, the designs are \textit{fundamentally different}. 
\THISWORK\ caches two pointers on the edge list in registers, without using shared memory. 
On the other hand, STMatch~\cite{stmatch22} only manages to cache the iterators and size of candidate arrays in shared memory, and still keep the content of candidate arrays in global memory (DRAM) due to its size. 
Its candidate arrays residing in global memory store its results for expensive vertex set operations, which are absent in the context of temporal motif mining.
As a result, \THISWORK\ can do caching in the fastest registers instead of using slower shared memory or global memory, like the solution in~\cite{stmatch22}. 

\section{Impact of $\delta$} \label{app:delta}
\begin{figure}[t]
\centerline{\includegraphics[width=\linewidth]{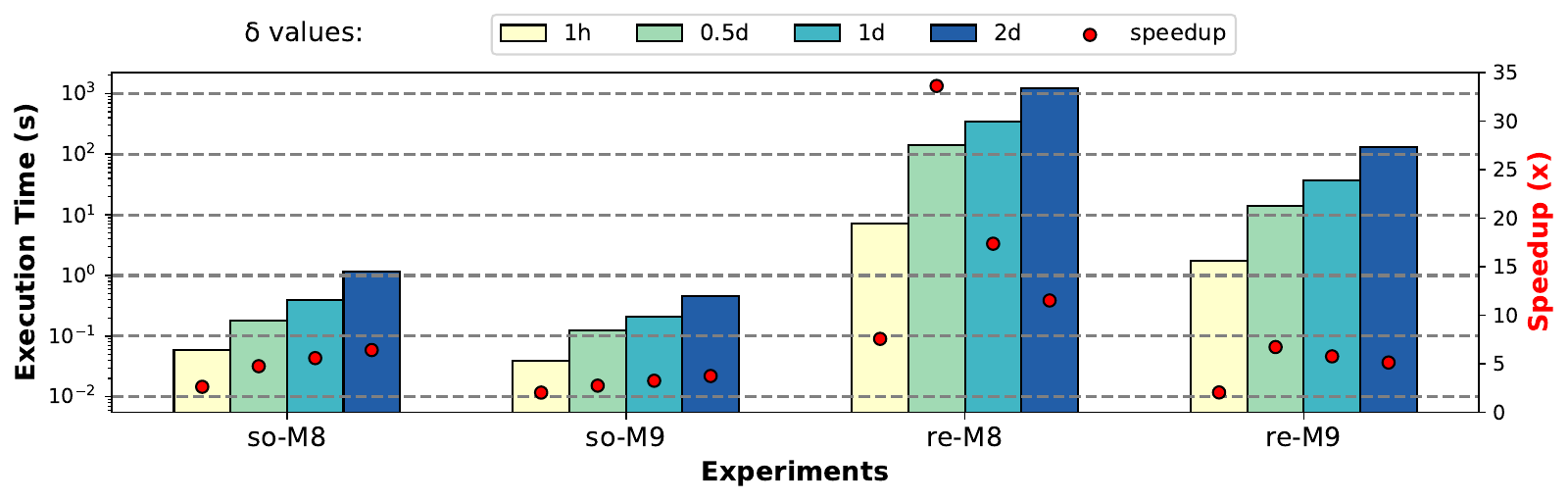}}
\caption{  Influence of $\delta$ on execution time and speedup over the baseline . \textllsm{so-M8} means mining \texttt{M8} on \textllsm{stackoverflow}. Other experiments follow the same naming convention. } 
\label{fig:delta}
\end{figure}

Figure \ref{fig:delta} illustrates how changing $\delta$ influences the execution time and speedup of \THISWORK\ over the baseline, mining two representative motifs on \textllsm{stackoverflow} and \textllsm{reddit-reply} as a case study. 
In practice, the execution time grows polynomially (close to linearly) with the time window $\delta$, which is better than the expectation from the worst case complexity analysis in Section 3.1 of the manuscript.
While \THISWORK\ always outperforms the baseline significantly, the speedup brought by \THISWORK\ varies for different $\delta$.
This is because $\delta$ determines the valid neighbors of vertices within search trees, thus changing the control flow behaviors and work distribution.

\section{Impact of Hyper-parameters} \label{app:hyper}

\begin{figure}[t]
\centerline{\includegraphics[width=\linewidth]{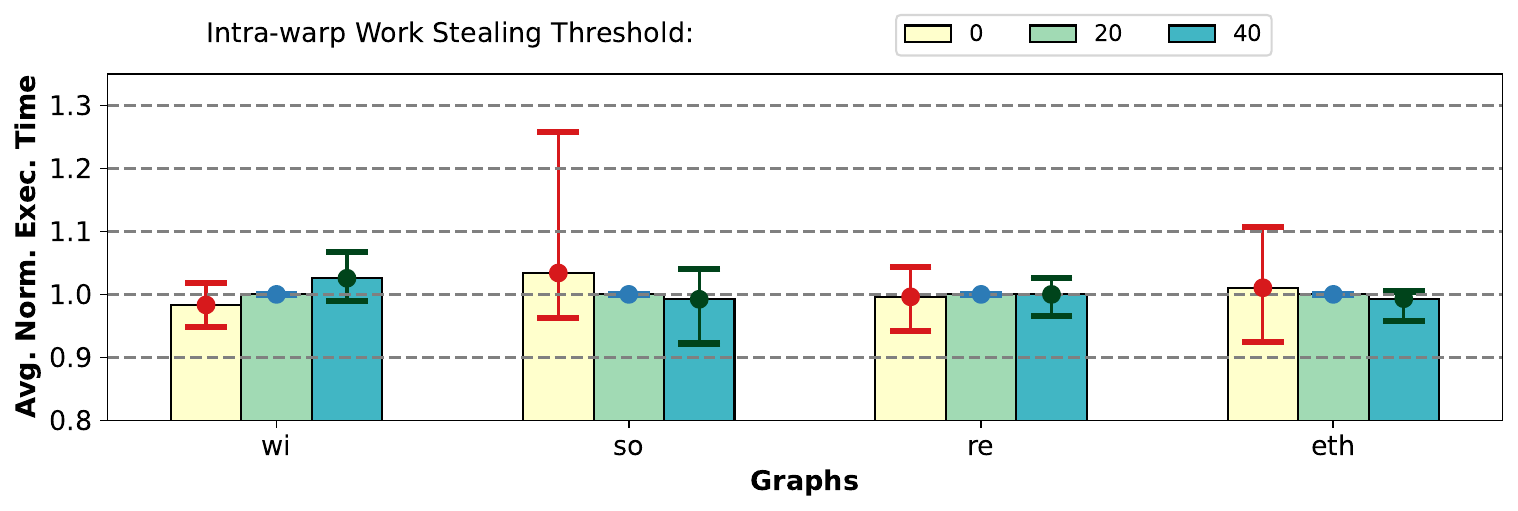}}
\caption{ Average execution time of the cases with different intra-warp work stealing threshold over the cases using default 20 iterations, all normalized to the cases with default settings.
The error bars show the maximum and minimum execution time mining different motifs in graphs. }
\label{fig:hyper-intra}
\end{figure}

\textbf{Choice of Intra-Warp Work Stealing Threshold.}
By default, intra-warp work stealing is enabled after 20 iterations to minimize the warp overhead that cannot benefit from load-balancing mechanisms. 
Such overhead can cause noticeable slowdowns when there is no severe work imbalance.
As shown in Figure \ref{fig:hyper-intra}, mining certain motifs on \texttt{stackoverflow} graph without using a threshold can consume 25\% more time than our default setting (the upper cap of \texttt{so-0}).
This 20-iteration threshold improves the speedup numbers in the cases where intra-warp work imbalance is not severe at the expense of a few percent slowdown in the highly-skewed cases, which benefit greatly from this optimization.
Thus, \THISWORK\ can more uniformly deliver high performance for all kinds of users' inputs. 
We use 20 iterations as the threshold, because we do not observe additional benefit using a larger value like 40, which is demonstrated in Figure \ref{fig:hyper-intra}.

\begin{figure}[t]
\centerline{\includegraphics[width=\linewidth]{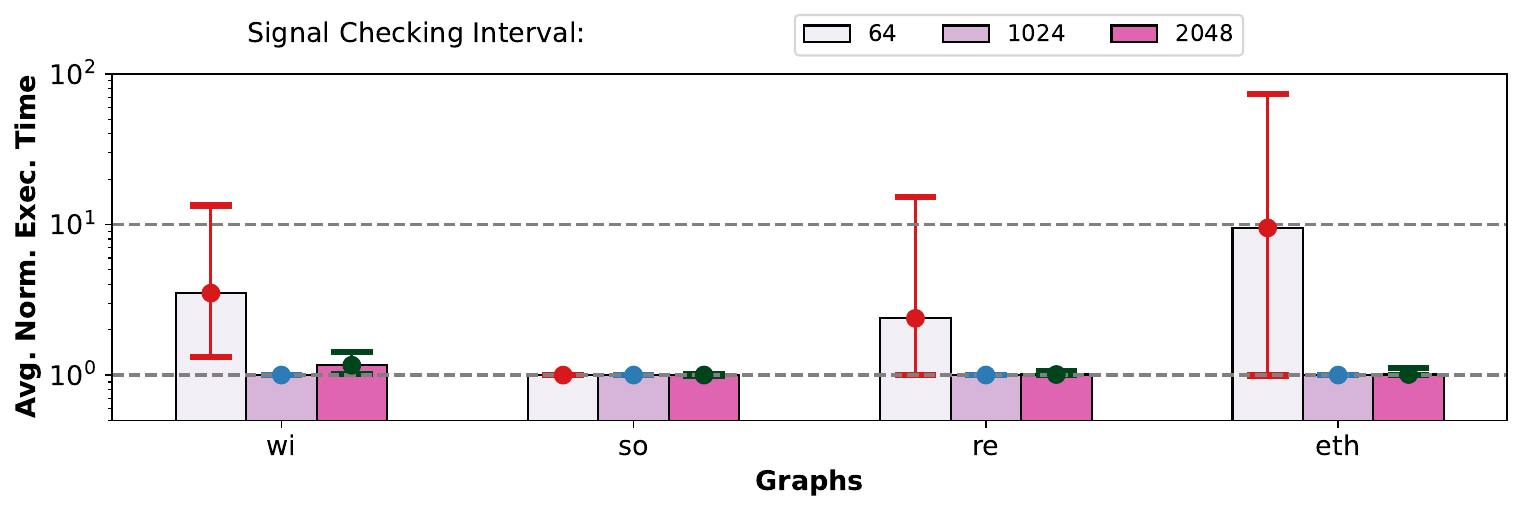}}
\caption{ Average execution time of the cases with different tail warp work redistribution signal checking intervals over the cases using default 1024 iterations, all normalized to the cases with default settings.
The error bars show the maximum and minimum execution time mining different motifs in graphs. 
}
\label{fig:hyper-it}
\end{figure}

\begin{figure}[t]
\centerline{\includegraphics[width=\linewidth]{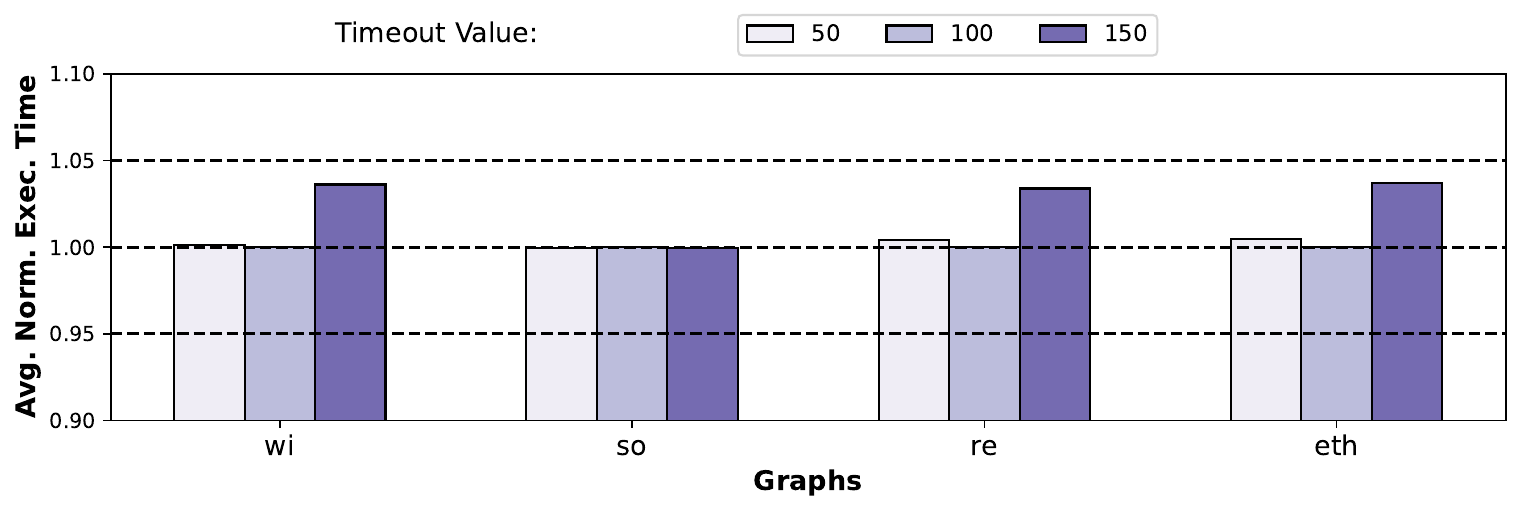}}
\caption{ Average execution time of the cases with different tail warp work redistribution timeout over the cases using default 100 ms, all normalized to the cases with default settings.
}
\label{fig:hyper-to}
\end{figure}

\noindent \textbf{Choice of Signal Checking Interval in Tail Warp Work Redistribution.}
To detect abortion signals, warps have to periodically check a signaling variable in global memory.
By default, \THISWORK\ uses 1024 iterations as the signal checking interval to amortize the memory accesses costs.
We experiment with different intervals for signal checking, as demonstrated in Figure \ref{fig:hyper-it}. 
If the signal checking intervals are set small (64 iterations), the system suffers from the overhead of frequent global memory accesses.
It does not slow down mining certain motifs due to their short search tree (especially the case for \texttt{stackoverflow}) which results in fewer signal checking.
However, on average, a small interval slows down the system by a few times for most of the graphs and is close to $100\times$ in the worst case.
Our choice, 1024 iterations, is a sweet point as no meaningful speedup is observed when we further increase this value.

\noindent \textbf{Choice of Timeout in Tail Warp Work Redistribution.}
We include a timeout value for this optimization to avoid too frequent recursive calls of tail warp work redistribution.
After receiving the abortion signals, warps will abort their execution after the timeout.
As shown in Figure~\ref{fig:hyper-to}, the overall performance of the system is not sensitive to this hyper-parameter.
The system performance will be roughly the same as long as we use reasonable values within the range shown in the figure.

\bibliographystyle{ACM-Reference-Format}
\bibliography{0_main}

\end{document}